\documentclass[amsmath,amssymb,
onecolumn,apm]{revtex4-2}

\usepackage{graphicx}
\usepackage{caption}

\begin{document}

\title{Observational constraints for cubic gravity theory based on third order contractions of the Riemann tensor}

\author{Mihai Marciu}
\email{mihai.marciu@drd.unibuc.ro}
 \author{Dana Maria Ioan}%
 \email{idana91@yahoo.com}
\author{Mihai Dragomir}%
 \email{mihai.dragomir@s.unibuc.ro}
\affiliation{ 
Faculty of Physics, University of Bucharest, Bucharest-Magurele, Romania
}

\date{\today}

\begin{abstract}
The paper studies different observational features in the case of a specific cubic gravity theory, based on third order contractions of the Riemann tensor. Considering viable cosmic chronometers data, baryon acoustic oscillations, and supernovae, we analyze the viability of such a theoretical model, obtaining specific constraints for different parameters in the current scenario. It is shown that the present extension of the $\Lambda$CDM cosmological model is compatible with recent data sets. The results indicate that the dark energy equation of state is exhibiting a  phantom regime in the near past in the case of the best fitted values, a behavior which is in agreement with various phenomenological studies.

\keywords{modified gravity \and dark energy}
\end{abstract}

\maketitle

\section{Introduction}
\label{intro}

\par 
In the present cosmological context \cite{Planck:2018vyg, WMAP:2010qai,Planck:2018jri, WMAP:2012fli}, the accelerated expansion \cite{Frieman:2008sn} plays a fundamental role, opening a variety of research directions in science and technology. Although the accelerated expansion has been discovered at the end of the last millennium \cite{SupernovaSearchTeam:1998fmf, SupernovaCosmologyProject:1998vns}, it remained an open problem \cite{Nojiri:2017ncd} in cosmological theories, driving the background dynamics of the Universe. This phenomenon, coined as dark energy, has been probed using different observational techniques \cite{SupernovaCosmologyProject:2011ycw, SupernovaCosmologyProject:2008ojh, Planck:2018nkj, Feng:2004ad, Weinberg:2013agg, SDSS:2004kqt, Allen:2011zs}. The simplest possible theoretical model is the $\Lambda$CDM model \cite{Copeland:2006wr}, a cosmological scenario which extends the Einstein--Hilbert action by adding a cosmological constant $\Lambda$, leading to a theoretical representation having a constant equation of state \cite{Joyce:2014kja}. The $\Lambda$CDM model can explain the dark energy phenomenon at the level of background dynamics \cite{Peebles:2002gy,Padmanabhan:2002ji}, representing a possible theoretical representation as an approximate effective theory. However, this model is not fully compatible to the evolution of the Universe \cite{Boylan-Kolchin:2011lmk, Perivolaropoulos:2021jda, Vagnozzi:2023nrq, Lopez-Corredoira:2017rqn, Pasten:2023rpc} and cannot explain several properties of the dark energy equation of state \cite{Wang:2022xdw, Zhao:2017cud, Upadhye:2004hh, DiValentino:2017zyq}. Consequently, various scalar tensor theories of gravitation have been proposed \cite{Capozziello:2011et, Nojiri:2010wj, Clifton:2011jh, Nojiri:2006ri, Tsujikawa:2010zza}, aiming towards a more complete theory of gravitation compatible to the evolution of our Universe \cite{Nojiri:2010wj}.
\par 
In these theories the $f(R)$ model appeared as a natural extension \cite{DeFelice:2010aj, Amendola:2006kh, Amarzguioui:2005zq}, based on the scalar curvature. This model can explain different aspects related to the evolution of the Universe, representing a possible direction in scalar tensor theories \cite{Sotiriou:2008rp, Nojiri:2006gh}. Later, various theories have appeared, based on different specific invariants \cite{Harko:2011kv, Cai:2015emx, Cognola:2006eg, Houndjo:2011tu, Wu:2010mn, Harko:2018gxr}. Another particular representation is related to the presence of scalar fields \cite{Carroll:2003st, Padmanabhan:2002cp, Ferreira:1997hj, Cai:2009zp, Deffayet:2010qz, Zimdahl:2001ar}, specific components which are also compatible to the accelerated expansion, driving the dark energy phenomenon. These fields can also be coupled to different specific invariants, in a non--minimally representation \cite{Nojiri:2005vv, Marciu:2019cpb, Marciu:2018oks, Bahamonde:2020vfj, Marciu:2020vve}. Hence, in the modified gravity theories, the accelerated expansion represents a curious phenomenon which is compatible to different representations \cite{Bahamonde:2017ize}, a compatibility acting on the fundamental geometry of space and time.
\par 
In the modified gravity theories a specific model has been developed recently, denoted as the Einsteinian cubic gravity \cite{Bueno:2016xff}. This model was introduced in an attempt of obtaining a more general theory of gravitation, by considering a linearization technique \cite{Bueno:2016xff}. Furthermore, a more generic gravity theory in a four dimensional space--time has been proposed  \cite{Erices:2019mkd}, a specific model which can explain the current accelerated expansion of the Universe. The dynamical analysis \cite{Quiros:2020eim} for such a theory has been investigated in Refs. \cite{Marciu:2020ysf, Quiros:2020uhr,Marciu:2021rdl}, analyzing specific properties of the background expansion. The inclusion of higher order invariants in the gravity theories has attracted a lot of attention in the recent years \cite{Bueno:2016ypa,Edelstein:2022xlb, Bueno:2019ltp,Caceres:2020jrf, Rudra:2022qbv, Bueno:2018xqc, BeltranJimenez:2020lee}. In Ref.~\cite{Giri:2021amc} the authors have considered different data sets, constraining the generic $f(P)$ cubic gravity theories using a machine learning approach. Also, the non--minimal coupling of the scalar fields with cubic invariants has been investigated recently \cite{Marciu:2022wzh, Marciu:2020ski, Marciu2023}. The inclusion of such higher order terms has applications in the inflationary epoch \cite{Arciniega:2018fxj,Edelstein:2020nhg, Arciniega:2019oxa, Arciniega:2018tnn}, in the study of black holes solutions \cite{Adair:2020vso, Bueno:2016lrh, Hennigar:2016gkm, Feng:2017tev}, with various ramifications in different aspects of cosmological theories \cite{Pookkillath:2020iqq, Hennigar:2018hza, Sardar:2021blt}.

\par 
In the present paper we shall further investigate the cubic gravity theory by adopting an observational perspective \cite{Bernardo:2021ynf}, considering different data sets, based on cosmic chronometers (CC), baryon acoustic oscillations (BAO), and supernovae (SNe). The approach will analyze the extension of the $\Lambda$CDM model towards a higher order gravity theory, considering a linear representation which takes into account a cubic invariant based on third order contractions of the Riemann tensor. This enables us to further examine the physical implications of such higher order terms in this geometric setup, obtaining viable constraints from the viewpoint of current observations.

\par 
The paper is organized as follows. In Sec.~\ref{sec:1} we briefly discuss the present extension of the $\Lambda$CDM model, presenting the corresponding fields equations which are obtained by varying the action with respect to the inverse metric. Then, in Sec.~\ref{sec:2} we discuss the basic ingredients used in order to apply the Markov-Chain-Monte-Carlo (MCMC) sampling, analyzing the obtained results. Lastly, in Sec.~\ref{sec:3} we give a short summary and the final concluding remarks.

\section{The cosmological model}
\label{sec:1}

\par 
In the following, we shall briefly present the cosmological model studied in the manuscript. We have considered a higher order extension of the $\Lambda$CDM model by embedding a specific cubic invariant in the corresponding action, based on third order contractions of the Riemann tensor. Hence, the theoretical scenario is described by the following action \cite{Erices:2019mkd, Bueno:2016xff}:

\begin{equation}
\label{actiune}
S=S_m+\int d^4x \sqrt{-g} \Bigg( \frac{1}{2}(R-2 \Lambda)+\alpha P\Bigg),\end{equation}
where $S_m$ represents the action corresponding to the matter component, $R$ the scalar curvature, $\alpha$ a constant parameter, and $P$ a higher order invariant which encodes the specific third order contractions of the Riemann tensor, 

\begin{multline}
P=\beta_1 R_{\mu\quad\nu}^{\quad\rho\quad\sigma}R_{ \rho\quad\sigma}^{\quad \gamma\quad\delta}R_{\gamma\quad\delta}^{\quad\mu\quad\nu}+\beta_2 R_{\mu\nu}^{\rho\sigma}R_{\rho\sigma}^{\gamma\delta}R_{\gamma\delta}^{\mu\nu}+\beta_3 R^{\sigma\gamma}R_{\mu\nu\rho\sigma}R_{\quad\quad\gamma}^{\mu\nu\rho}+\beta_4 R R_{\mu\nu\rho\sigma}R^{\mu\nu\rho\sigma}+\beta_5 R_{\mu\nu\rho\sigma}R^{\mu\rho}R^{\nu\sigma}
\\+\beta_6 R_{\mu}^{\nu}R_{\nu}^{\rho}R_{\rho}^{\mu}+\beta_7 R_{\mu\nu}R^{\mu\nu}R+\beta_8 R^3.
\end{multline}
In the previous relation $\beta_i$ ($i={1,..,8}$) are constant coefficients specific to the cubic gravity theory. If we take into account the following relations \cite{Erices:2019mkd, Bueno:2016xff}, 

\begin{equation}
\beta_7=\frac{1}{12}\big[3\beta_1-24\beta_2-16\beta_3-48\beta_4-5\beta_5-9\beta_6\big],
\end{equation}

\begin{equation}
\beta_8=\frac{1}{72}\big[-6\beta_1+36\beta_2+22\beta_3+64\beta_4+3\beta_5+9\beta_6\big],
\end{equation}

\begin{equation}
\beta_6=4\beta_2+2\beta_3+8\beta_4+\beta_5,
\end{equation}

we obtain the specific cubic gravity theory which encodes third order contractions of the Riemann tensor. In this case we use the specific redefinition, 

\begin{equation}
\bar{\beta}=(-\beta_1+4\beta_2+2\beta_3+8\beta_4),
\end{equation}

obtaining the following representation for the higher order term \cite{Erices:2019mkd, Bueno:2016xff}:

\begin{equation}
\label{PP}
P=6\bar{\beta}H^4 (2H^2+3\dot{H}),
\end{equation}

in the case of a homogeneous and isotropic Universe characterized by the Robertson--Walker metric, 

\begin{equation}
\label{metrica}
ds^2=-dt^2+a^2(t) \delta_{ik}dx^i dx^k,
\end{equation}
with $a(t)$ the cosmic scale factor. If we consider the variation of the previous action \eqref{actiune} with respect to the inverse metric, we obtain the following modified Friedmann relations \cite{Erices:2019mkd, Bueno:2016xff}:

\begin{equation}
\label{eqfr1}
3H^2=\rho_m+\rho_{g},
\end{equation}

\begin{equation}
\label{eqfr2}
3H^2+2\dot{H}=-p_m-p_{g},
\end{equation}

with $\rho_g$ the density of the geometrical dark energy component, and $p_g$ the corresponding pressure, 

\begin{equation}
    \rho_g=6 \alpha \bar{\beta} H^6 +\Lambda
\end{equation}

\begin{equation}
    p_g=-6 \alpha \bar{\beta} H^4 (H^2+2 \dot{H})-\Lambda.
\end{equation}
In what follows we can redefine the specific cubic constant, $\beta=\alpha \bar{\beta}$, obtaining  simpler expressions for the previous relations. Finally, we can write the associated density parameters as usual, 

\begin{equation}
    \Omega_m=\frac{\rho_m}{3 H^2},
\end{equation}

\begin{equation}
    \Omega_g=\frac{\rho_g}{3 H^2},
\end{equation}

satisfying the corresponding constraint,

\begin{equation}
   \Omega_m+\Omega_g=1.
\end{equation}

\section{Observational properties for the cubic gravity theory}
\label{sec:2}

\par 
In order to study the higher order gravity theory from an observational perspective, we need to express the dynamical equations in a comprehensible manner. Combining the equations \eqref{eqfr1} and \eqref{eqfr2}, we arrive at the following expression:
\begin{equation}
\label{eqz1}
    -3 H^2+6 \beta H^6+\Lambda=\dot{H}(2-12 \beta H^4).
\end{equation}
Next, we introduce the redshift variable defined as:
\begin{equation}
    z(t)=\frac{1}{a(t)}-1,
\end{equation}
changing the dependence from the cosmic time to redshift. In this case, the time derivative of the Hubble parameter can be expressed as follows:
\begin{equation}
    \dot{H}=(-H)H'(1+z),
\end{equation}
where the prime indicates the derivative with respect to the redshift variable. The equation \eqref{eqz1} takes the following form:
\begin{equation}
    3 H^2-6 \beta H^6-\Lambda=H H' (1+z)(2-12 \beta H^4).
\end{equation}
Next, we introduce the dimensionless Hubble parameter defined as:
\begin{equation}
    E(z)=\frac{H(z)}{H_0},
\end{equation}
expressing the evolution of the cosmological system in an autonomous manner,
\begin{equation}
    E'=\frac{3 H_0^2E^2-6 \beta E^6 H_0^6-\Lambda}{E H_0^2(1+z)(2-12 \beta E^4 H_0^4)}.
\end{equation}
We now introduce the density parameter associated to the cosmological constant $\Lambda$ at the present time, 
\begin{equation}
\Omega_{\Lambda}=\frac{\Lambda}{3 H_0^2},
\end{equation}
the density parameter for the matter component,
\begin{equation}
   \Omega_{m0}=\frac{\rho_m(z=0)}{3 H_0^2}, 
\end{equation}
and the density parameter corresponding to the higher order corrections of the $\Lambda$CDM model,
\begin{equation}
   \Omega_{c0}=2 \beta H_0^4, 
\end{equation}
satisfying the following constraint at $z=0$:
\begin{equation}
  \Omega_{m0}+\Omega_{\Lambda}+ \Omega_{c0}=1. 
\end{equation}
Hence, the evolution of the dynamical cosmological system can be expressed as follows:
\begin{equation}
\label{eqff}
    E'=\frac{3 E^2-6 \beta E^6 H_0^4-3\Omega_{\Lambda}}{2 E (1+z)(1-6 \beta E^4 H_0^4)}.
\end{equation}
In order to emulate the evolution of the cosmological model we solve the ordinary differential equation \eqref{eqff} taking into account that at the present time ($z=0$) we have the initial condition, 
\begin{equation}
    E(z=0)=\frac{H(z=0)}{H_0}=1.
\end{equation}
As can be noted, the corresponding evolution \eqref{eqff} depends on the present value of the Hubble parameter ($H_0$), the current density parameter associated to the cosmological constant $\Lambda$, and the $\beta$ parameter, a constant expressing the strength of the higher order corrections based on third order contractions of the Riemann tensor.
\par
\begin{figure}[t]
  \includegraphics[width=17cm]{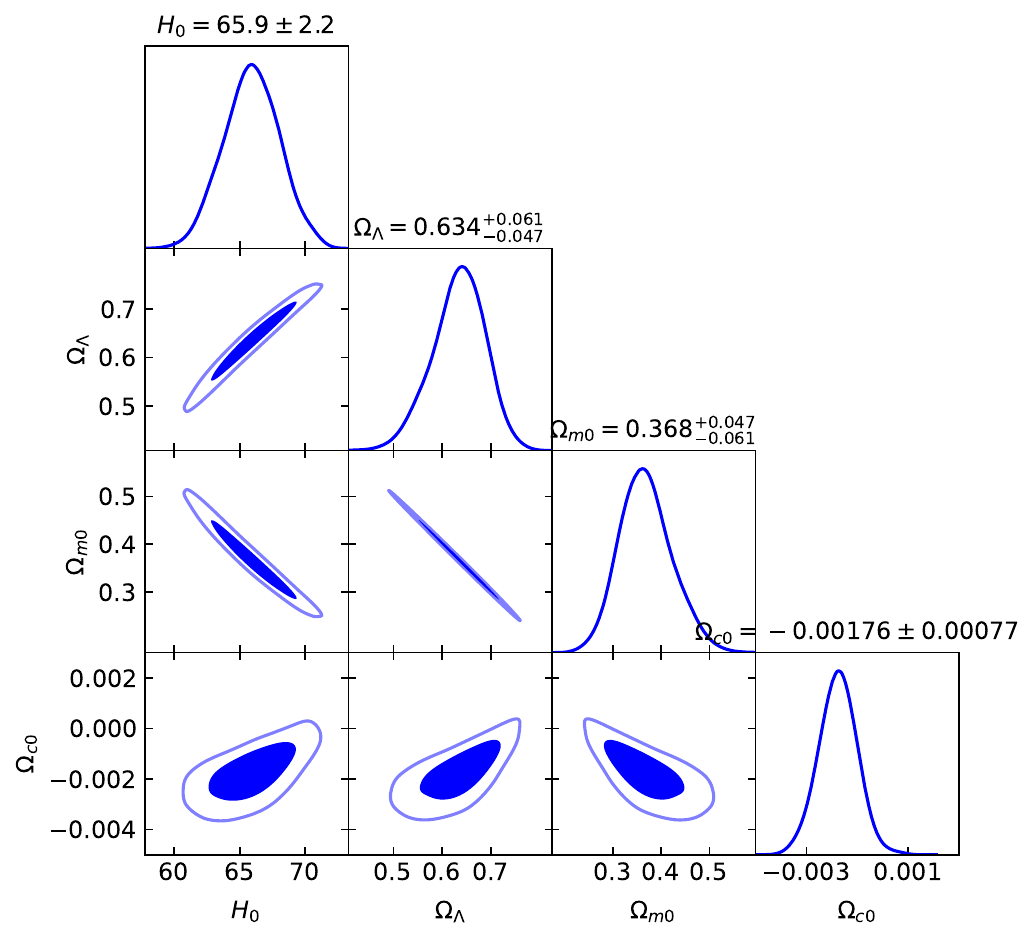} 
\caption{The posterior distributions for the relevant cosmological parameters in the case of cosmic chronometers and baryon acoustic oscillations observations.}
\label{fig:b1}       % Give a unique label
\end{figure}

\begin{figure}[t]
\includegraphics[width=6cm]{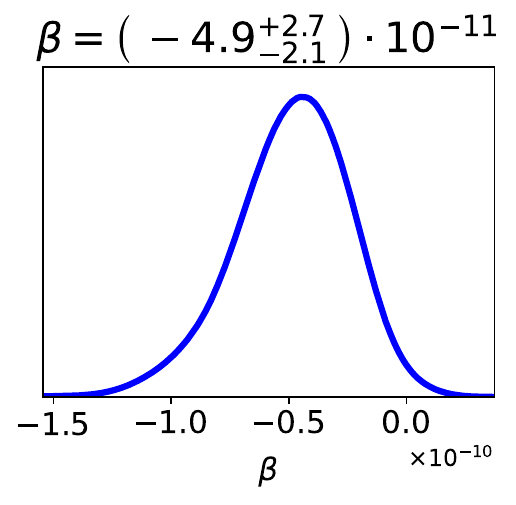}  
\caption{One dimensional posterior distribution for the $\beta$ parameter in the case of CC +  BAO observations.}
\label{fig:b2}       % Give a unique label
\end{figure}

\begin{figure}[t]
\includegraphics[width=10cm]{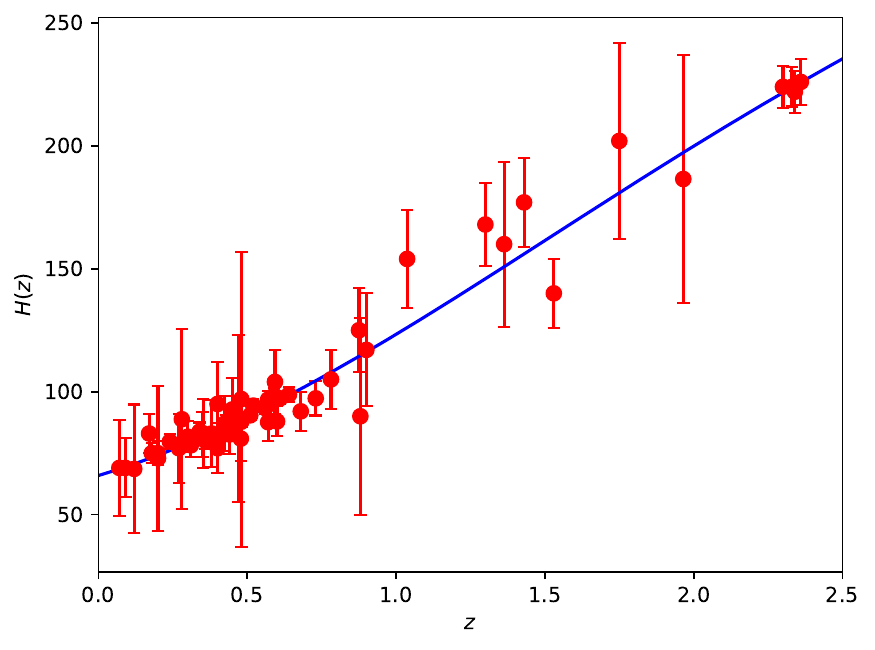}  
\caption{The variation of the Hubble parameter in the case of CC +  BAO observations.}
\label{fig:b3}       % Give a unique label
\end{figure}

\begin{figure}[t]
\includegraphics[width=10cm]{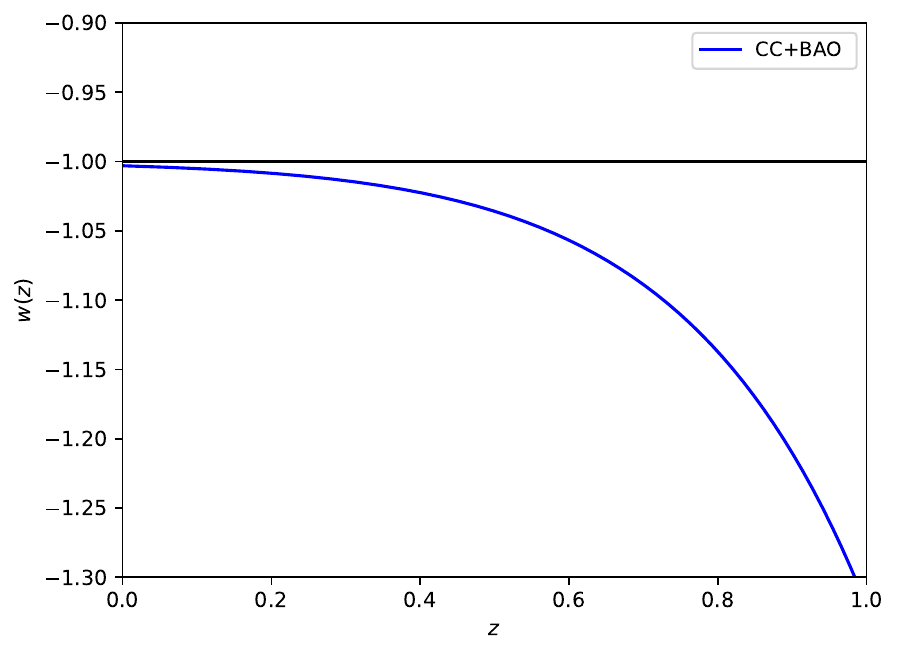}  
\caption{The evolution of the dark energy equation of state for the best fitted values in the case of CC +  BAO observations.}
\label{fig:b4}       % Give a unique label
\end{figure}

\begin{figure}[t]
  \includegraphics[width=17cm]{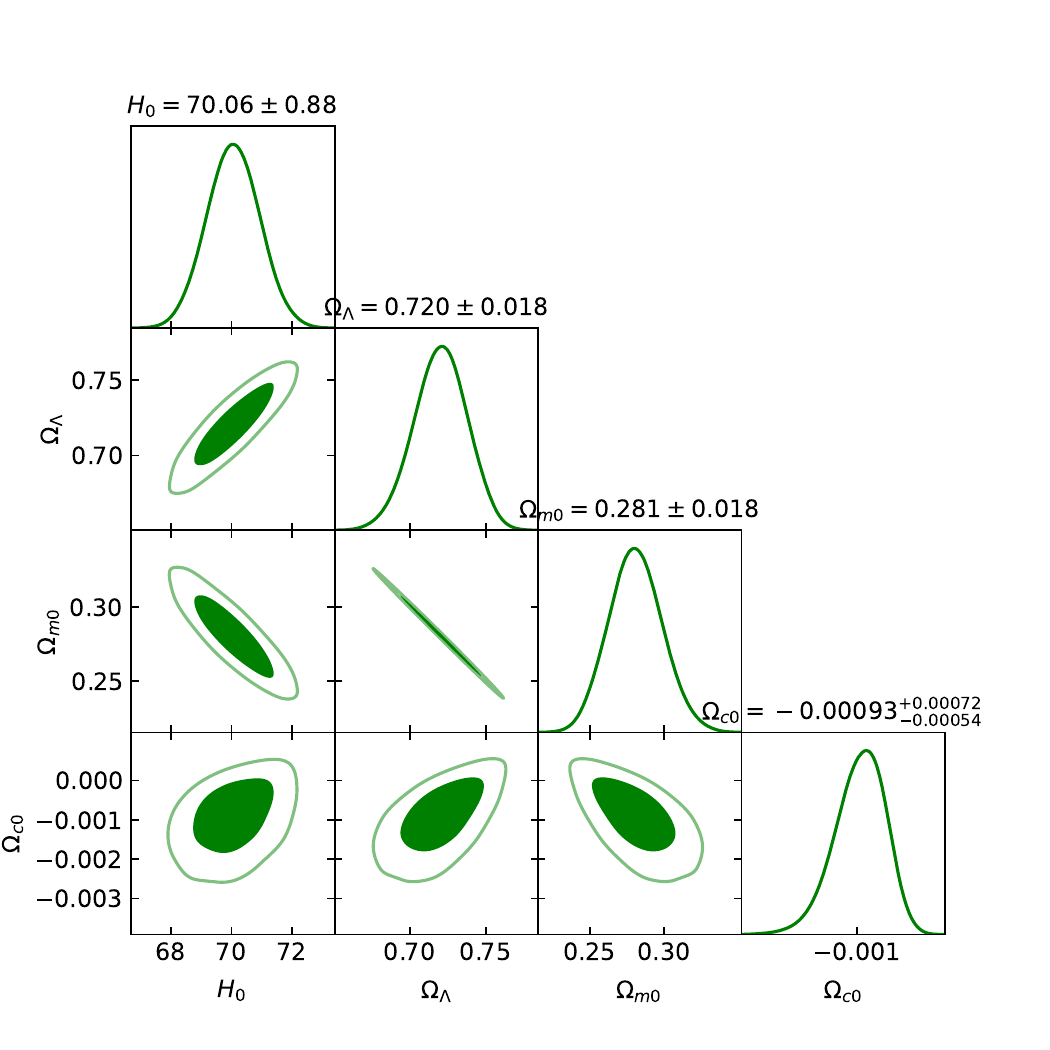} 
\caption{The posterior distributions for the relevant cosmological parameters in the case of CC +  BAO + SNe observations.}
\label{fig:b5}       % Give a unique label
\end{figure}

\begin{figure}[t]
\includegraphics[width=6cm]{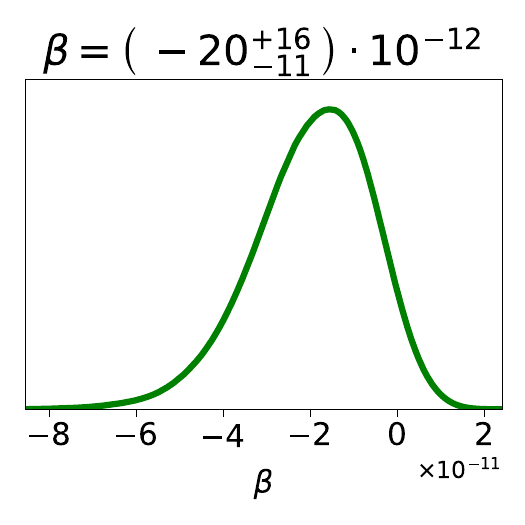}  
\caption{One dimensional posterior distribution for the $\beta$ parameter in the case of CC +  BAO + SNe observations. }
\label{fig:b6}       % Give a unique label
\end{figure}

\begin{figure}[t]
\includegraphics[width=6cm]{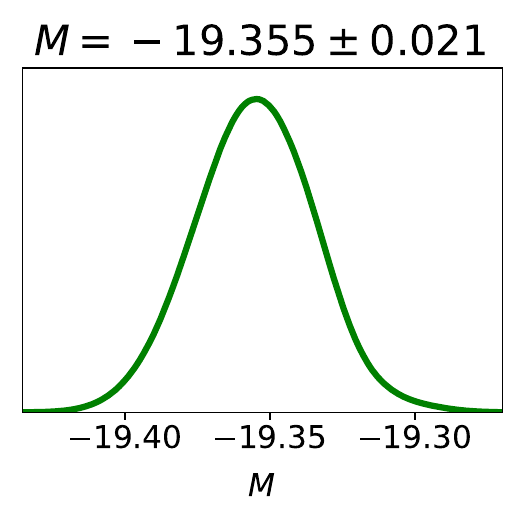}  
\caption{One dimensional posterior distribution for the $M$ parameter in the case of CC +  BAO + SNe observations. }
\label{fig:b7}       % Give a unique label
\end{figure}

\begin{figure}[t]
\includegraphics[width=10cm]{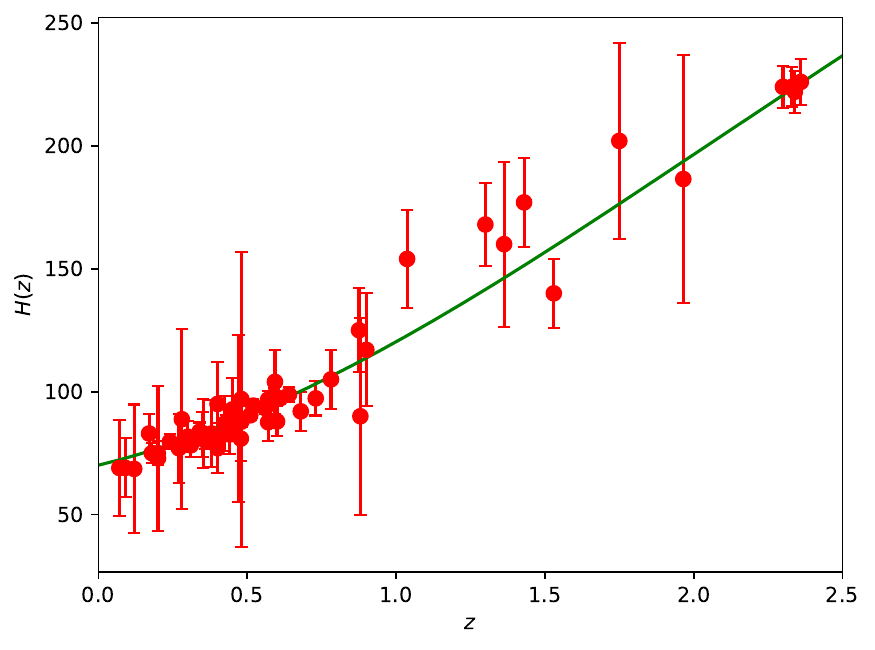}  
\caption{The variation of the Hubble parameter in the case of CC +  BAO + SNe observations.}
\label{fig:b8}       % Give a unique label
\end{figure}

\begin{figure}[t]
\includegraphics[width=10cm]{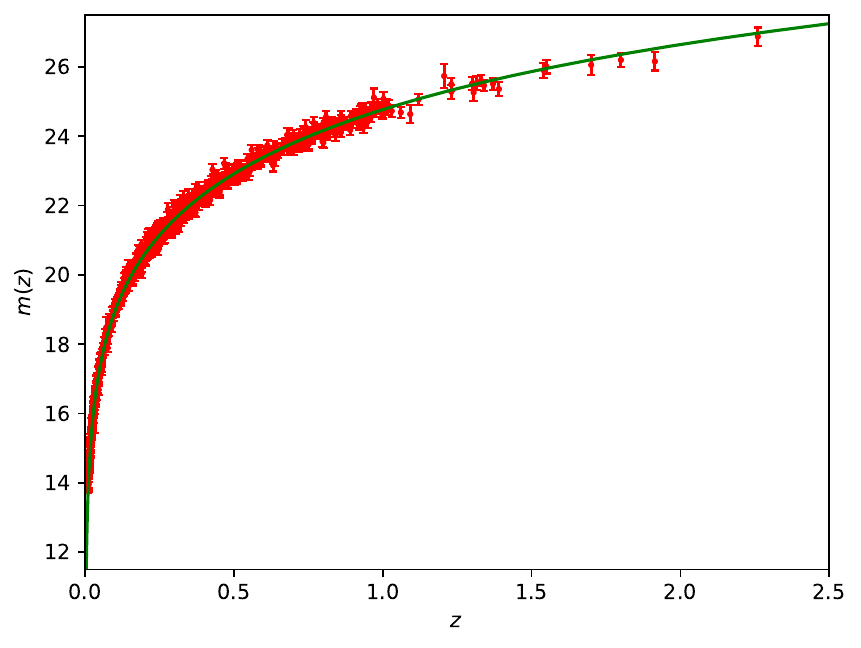}  
\caption{The variation of the apparent magnitude of the supernovae in the case of the best fitted values for CC +  BAO + SNe observations.}
\label{fig:b9}       % Give a unique label
\end{figure}

\begin{figure}[t]
\includegraphics[width=10cm]{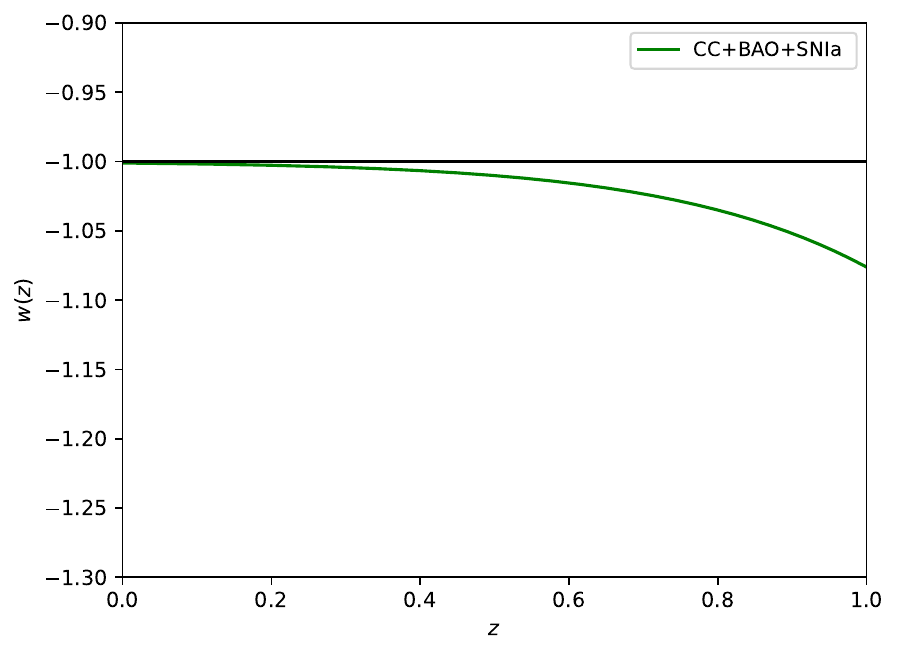}  
\caption{The dark energy equation of state for the best fitted values in the case of CC +  BAO + SNe observations.}
\label{fig:b10}       % Give a unique label
\end{figure}

 
In order to study the present cosmological model, we have considered specific observations based on cosmic chronometers (CC), baryon acoustic oscillations (BAO), and Ia supernovae (SNe). This allows us to further constrain the corresponding free parameters in the equation \eqref{eqff}, $(H_0, \Omega_{\Lambda}, \beta)$. For the observational analysis, we have considered data from CC, BAO, and SNe \cite{Bernardo:2021ynf}, obtaining possible viable constraints associated to the higher order corrections due to the specific contractions of the Riemann tensor. In our analysis we have defined the associated chi-squared functions. For the CC + BAO data the chi-square has the following form:
\begin{equation}    \chi^2_{CC+BAO}=\sum_{i=1}^{31(CC)+26(BAO)}\left(\frac{H_o(z_i)-H_m(z_i)}{\sigma_{i}}\right)^2,
\end{equation}
where $H_o$ represents the observed value of the Hubble parameter having the corresponding uncertainty $\sigma_i$ at the redshift $z_i$, while $H_m$ denotes the value obtained from the present cosmological model (obtained by numerical evolution of the eq. \eqref{eqff}). The results of the Markov chain Monte-Carlo sampling for the CC + BAO data have been presented in Figs.~\ref{fig:b1}--\ref{fig:b4}. In order to obtain these figures, we have used the Cobaya package \cite{Bernardo:2021ynf,Torrado:2020dgo} considering the following value for the Gelman-Rubin criterion $R-1=0.001$. Secondly, for the CC + BAO + SNe dataset we have further defined a more general chi--square function, 
\begin{equation}    \chi^2_{CC+BAO+SNe}=\chi^2_{CC+BAO}+\chi^2_{SNe},
\end{equation}
where 
\begin{equation}    
\chi^2_{SNe}=\sum_{z_1}\sum_{z_2} \left(m_o(z_1)-m_m(z_1))\right)C^{-1}(z_1,z_2)(m_o(z_2)-m_m(z_2)),
\end{equation}
with $m$ the apparent magnitude having the specific covariance matrix $C(z_1,z_2)$. As in the previous case, the subscript (o) is associated to the observations, while (m) corresponds to the values obtained using the present cosmological model \cite{Bernardo:2021ynf,Torrado:2020dgo}. For the second sampling, in the case of CC + BAO + SNe data set, we have used 1048 supernovae type Ia events from the Pantheon data \cite{Pan-STARRS1:2017jku}. The results of the second analysis are presented in Figs.~\ref{fig:b5}--\ref{fig:b10}. In what follows we shall briefly discuss the obtained results specific for the higher order contractions of the Riemann tensor. In Fig.~\ref{fig:b1} we present the posterior distributions and the corresponding confidence intervals for the higher order corrections of the $\Lambda$CDM model by taking into account data from cosmic chronometers and baryon acoustic oscillations. In this case we note that the density parameter associated to the matter content $\Omega_{m0}$ has a higher value, while the density parameter corresponding to the higher order corrections $\Omega_{c0}$ is negative and $\propto 10^{-3}$. In Fig.~\ref{fig:b2} we present one dimensional distribution associated to the $\beta$ parameter, observing that $\beta \propto 10^{-11}$. The variation of the best fitted value corresponding to the Hubble parameter is presented in Fig.~\ref{fig:b3}, while the corresponding equation of state for the geometrical dark energy component is displayed in Fig.~\ref{fig:b4}. From the evolution of the geometrical dark energy equation of state ($w_g=p_g / \rho_g$) we observe that at late times the plot follows closely the $\Lambda$CDM model, while at early time the deviation corresponds to a super--accelerated regime. The best fitted values obtained suggest that the main source of cosmic acceleration is related to the cosmological constant, while the higher order corrections based on the third order contractions of the Riemann tensor represent the origin of the super--accelerated regime, explaining the early times dynamics and the non--negligible deviations from the phantom divide line in the near past.
\par 
For the CC + BAO + SNe observations we have presented the corresponding confidence intervals in Fig.~\ref{fig:b5}, observing that the value of the present Hubble parameter is higher in this case (as compared to CC + BAO data set), having a lower value of the matter density parameter ($\Omega_{m0}$) at the present time. The results indicate that the density parameter corresponding to the higher order contractions of the Riemann tensor is negative and non--negligible, closely proportional to $\Omega_{c0} \propto 10^{-3}$. The result is almost consistent with the CC + BAO data set, suggesting a compatibility between the two data sets. Furthermore, from Fig.~\ref{fig:b6} we see that the value of the $\beta$ parameter is $\propto 10^{-11}$, a result consistent with the previous data set. In Fig.~\ref{fig:b7} we have displayed the associated one dimensional posterior distribution for the absolute magnitude. Next, in Figs.~\ref{fig:b8}--\ref{fig:b9} we have showed the compatibility of the best fitted results with respect to observations, for the variation of the associated Hubble parameter and the corresponding apparent magnitude. Lastly, in Fig.~\ref{fig:b10} we have displayed the evolution of the dark energy equation of state, observing the presence of the super--accelerated regime in the near past. The results are consistent with the CC + BAO data sets, noting that for the Pantheon data the regime is less super--accelerated. The obtained values for the two data sets are also displayed in Table \ref{tabel1}, summarizing the present results.

\begin{center}
\captionof{table}{The confidence intervals for the higher order gravity theory based on cubic contractions of the Riemann tensor.}
\begin{tabular}{| c | c | c | c | c | c |}
\hline
 Data set & $H_0$ & $\Omega_{\Lambda}$ & $\Omega_{m0}$ & $\Omega_{c0}$ & $\beta$ \\ 
 \hline
 CC + BAO & $65.9\pm 2.2$ & $0.634^{+0.061}_{-0.047}$ & $0.368^{+0.047}_{-0.061}$& $-0.00176 \pm 0.00077$ & $-4.9^{+2.7}_{-2.1} \cdot 10^{-11}$\\
 \hline
 CC + BAO  + SNe & $70.06\pm 0.88$ & $0.720 \pm 0.018$ & $0.281 \pm 0.018$& $-0.00093^{+0.00072}_{-0.00054}$ & $-2.0^{+1.6}_{-1.1} \cdot 10^{-11}$\\
 \hline   
\end{tabular}
\label{tabel1}
\end{center}

\begin{center}
\captionof{table}{The confidence intervals for the $\Lambda$CDM model.}
\begin{tabular}{| c | c | c | c  |}
\hline
 Data set & $H_0$ & $\Omega_{\Lambda}$ & $\Omega_{m0}$ \\ 
 \hline
 CC + BAO & $69.75 \pm 1.10$ & $0.72 \pm 0.018$ & $0.27 \pm 0.018$ \\
 \hline
  CC + BAO + SNe & $70.59\pm 0.78$ & $0.73 \pm 0.012$ & $0.26 \pm 0.012$ \\
 \hline   
\end{tabular}
\label{tabel2}
\end{center}

\par 
We have also calculated the specific values of the Akaike information criterion (AIC),
\begin{equation}
    AIC=2 k -2 Log(L_{max}),
\end{equation}
and Bayesian information criterion (BIC),
\begin{equation}
    BIC=k Log (N) -2 Log(L_{max}).
\end{equation}
In these formulas $L_{max}$ denotes the maximum likelihood, $N$ is associated to data, while $k$ corresponds to the number of parameters in the sampling. The results are shown in Table \ref{tabel3}, where the displayed results are obtained by subtracting the corresponding value from the result specific to the $\Lambda$CDM model. The results show that from the viewpoint of $\chi^2$ statistic and the Akaike information criterion the present model is preferred with respect to the $\Lambda$CDM model. This result is valid for the two data sets, CC + BAO, and CC + BAO + SNe. However, considering the Bayesian information criterion, the situation is different. For CC + BAO data, the model is slight better than the $\Lambda$CDM model, while for the Pantheon systematic the $\Lambda$CDM model is still favored. As can be seen, those two data sets are in tension with each other, affecting the obtained results. Hence, the present model is favored by the data coming from cosmic chronometers and baryon acoustic oscillations. The extension of the $\Lambda$CDM model towards the higher order gravity theories based on third order contractions of the Riemann tensor represents a possible and viable scenario which can explain the accelerated expansion, with the main source of cosmic acceleration driven by the cosmological constant.    

\begin{center}
\captionof{table}{The AIC and BIC values corresponding to the present cosmological scenario.}
\begin{tabular}{| c | c | c | c  |}
\hline
 Data set & $\Delta \chi^2$ & $\Delta AIC$ & $\Delta BIC$ \\ 
 \hline
 CC + BAO & $4.43$ & $2.43$ & $0.39$ \\
 \hline
  CC + BAO + SNe & $3.23$ & $1.23$ & $-3.76$ \\
 \hline   
\end{tabular}
\label{tabel3}
\end{center}

\section{Summary and Conclusions}
\label{sec:3}
\par 
In the present paper we have studied a modified gravity theory which extends the fundamental $\Lambda$CDM model by embedding viable geometrical constituents based on third order contractions of the Riemann tensor. This cosmological model represents a novel approach in the modified gravity theories, taking into account a specific invariant which leads to second order equations in the Friedmann relations. After briefly describing the proposed action and the corresponding modified Friedmann relations, we have introduced the dimensionless Hubble parameter, expressing the evolution of the dynamical system in a proper manner. The present study is devoted to the analysis of the particular extension of the $\Lambda$CDM model from an observational perspective, investigating specific viable constraints for the cubic gravity model in the linear order. The analysis takes into account data from cosmic chronometers, baryon acoustic oscillations, and supernovae of Ia type as the Pantheon data set. After expressing the corresponding $\chi^2$ specific functions, we have used the Cobaya package for the Markov chain Monte-Carlo sampling. The results of our analysis have been discussed in the previous sections for the CC + BAO and CC + BAO + SNe data sets.
\par 
From the statistical analysis it is clear that a realistic modified gravity theory should go beyond the $\Lambda$CDM model, the simplest cosmological scenario. In our approach we have considered higher order terms in the proposed action, based on third order contractions of the Riemann tensor. In the previous section we have seen that in the case of CC + BAO and CC + BAO + SNe data sets, the inclusion of such higher order terms is mildly preferred and fully compatible to recent cosmological data. However, the problem of $H_0$ tension is still present in the current analysis, due to the deviations of the corresponding observables.
\par 
In the case of CC + BAO data set we have observed that the values corresponding to the best fitted parameters are comparable with the specific values for the $\Lambda$CDM model. Hence, the value of the present Hubble parameter $H_0$ is quite lower for the CC + BAO data set, while the matter density parameter at the present time $\Omega_{m0}$ has a higher value. The density parameter associated to the higher order contractions of the Riemann tensor $\Omega_{c0}$ is negative, with the $\beta$ parameter negative and proportional to $\beta \propto 10^{-11}$. All of these suggests that the present higher order extension of the $\Lambda$CDM model acts in the early evolution of the Universe, diminishing the accelerated expansion in the early matter era. It is also interesting to note that the dark energy equation of state associated to the geometrical component has a phantom behavior, without the inclusion of the scalar fields with a negative kinetic energy.
\par
The results for the CC + BAO + SNe data sets show that the inclusion of the higher order contractions of the Riemann tensor leads to a slightly lower value for the Hubble parameter at the present time (compared to the $\Lambda$CDM model). Moreover, the best fitted value for the matter density parameter $\Omega_{m0}$ is higher with respect to the corresponding value specific for the $\Lambda$CDM model, suggesting that the inclusion of the cubic components slightly moderate the accelerated expansion in the near past. The present analysis has also considered the Akaike information criterion and the Bayesian information criterion for the previous mentioned data sets. We note that for the Akaike information criterion the inclusion of cubic contractions of the Riemann tensor is slightly preferred for the two data sets. However, in the case of Bayesian information criterion the results are not conclusive. For the CC + BAO data, the higher order contractions of the Riemann tensor are very slightly preferred, while for the Pantheon data the $\Lambda$CDM model is still favored. We mention that for the CC + BAO + SNe data sets the geometrical dark energy component has a phantom behavior, acting bellow the cosmological constant boundary. However, the inclusion of the Pantheon data moderate the cosmic expansion, with the corresponding variation less accelerated. In this case the value of the density parameter associated to the higher order contractions of the Riemann tensor $\Omega_{c0}$ is still negative and comparable to the previous values.
\par 
The present cosmological model can be further extended in a variety of approaches. For example, one can consider an extension of the corresponding action beyond the linear order, discussing the full compatibility of the $f(P)$ gravity theory with different observations. Moreover, the statistical analysis can also take into account various data sets, defining a more stringent $\chi^2$ function. This would allow us to obtain a more realistic theoretical model which includes higher order terms fully compatible to recent cosmological data. All of these questions are open and left for a different investigation in the near future.

\section{data availability}
\label{sec:6}
Data sharing not applicable to this article as no datasets were generated or analysed during the current study.

\section{Acknowledgements}
The authors would like to thank Prof. Dr. Virgil Baran for continuous support and many fruitful discussions. For the computational part of the present project we have used the new established cluster from the Faculty of Physics, University of Bucharest. The work was partially supported by the project 41PFE/30.12.2021, financed by the Ministry of Research, Innovation and Digitalization through Program 1 – Development of the National R$\and$D System, Subprogram 1.2. Institutional performance – Financing projects for excellence in RDI. Different computations have been done in Wolfram Mathematica \cite{Mathematica} and xAct \cite{xact}. 

\bibliography{sorsamp}% Produces the bibliography via BibTeX.

\providecommand{\noopsort}[1]{}\providecommand{\singleletter}[1]{#1}%
\begin{thebibliography}{89}%
\makeatletter
\providecommand \@ifxundefined [1]{%
 \@ifx{#1\undefined}
}%
\providecommand \@ifnum [1]{%
 \ifnum #1\expandafter \@firstoftwo
 \else \expandafter \@secondoftwo
 \fi
}%
\providecommand \@ifx [1]{%
 \ifx #1\expandafter \@firstoftwo
 \else \expandafter \@secondoftwo
 \fi
}%
\providecommand \natexlab [1]{#1}%
\providecommand \enquote  [1]{``#1''}%
\providecommand \bibnamefont  [1]{#1}%
\providecommand \bibfnamefont [1]{#1}%
\providecommand \citenamefont [1]{#1}%
\providecommand \href@noop [0]{\@secondoftwo}%
\providecommand \href [0]{\begingroup \@sanitize@url \@href}%
\providecommand \@href[1]{\@@startlink{#1}\@@href}%
\providecommand \@@href[1]{\endgroup#1\@@endlink}%
\providecommand \@sanitize@url [0]{\catcode `\\12\catcode `\$12\catcode `\&12\catcode `\#12\catcode `\^12\catcode `\_12\catcode `\%12\relax}%
\providecommand \@@startlink[1]{}%
\providecommand \@@endlink[0]{}%
\providecommand \url  [0]{\begingroup\@sanitize@url \@url }%
\providecommand \@url [1]{\endgroup\@href {#1}{\urlprefix }}%
\providecommand \urlprefix  [0]{URL }%
\providecommand \Eprint [0]{\href }%
\providecommand \doibase [0]{https://doi.org/}%
\providecommand \selectlanguage [0]{\@gobble}%
\providecommand \bibinfo  [0]{\@secondoftwo}%
\providecommand \bibfield  [0]{\@secondoftwo}%
\providecommand \translation [1]{[#1]}%
\providecommand \BibitemOpen [0]{}%
\providecommand \bibitemStop [0]{}%
\providecommand \bibitemNoStop [0]{.\EOS\space}%
\providecommand \EOS [0]{\spacefactor3000\relax}%
\providecommand \BibitemShut  [1]{\csname bibitem#1\endcsname}%
\let\auto@bib@innerbib\@empty
%</preamble>
\bibitem [{\citenamefont {Aghanim}\ \emph {et~al.}(2020{\natexlab{a}})\citenamefont {Aghanim} \emph {et~al.}}]{Planck:2018vyg}%
  \BibitemOpen
  \bibfield  {author} {\bibinfo {author} {\bibfnamefont {N.}~\bibnamefont {Aghanim}} \emph {et~al.} (\bibinfo {collaboration} {Planck}),\ }\bibfield  {title} {\bibinfo {title} {{Planck 2018 results. VI. Cosmological parameters}},\ }\href {https://doi.org/10.1051/0004-6361/201833910} {\bibfield  {journal} {\bibinfo  {journal} {Astron. Astrophys.}\ }\textbf {\bibinfo {volume} {641}},\ \bibinfo {pages} {A6} (\bibinfo {year} {2020}{\natexlab{a}})},\ \bibinfo {note} {[Erratum: Astron.Astrophys. 652, C4 (2021)]},\ \Eprint {https://arxiv.org/abs/1807.06209} {arXiv:1807.06209 [astro-ph.CO]} \BibitemShut {NoStop}%
\bibitem [{\citenamefont {Komatsu}\ \emph {et~al.}(2011)\citenamefont {Komatsu} \emph {et~al.}}]{WMAP:2010qai}%
  \BibitemOpen
  \bibfield  {author} {\bibinfo {author} {\bibfnamefont {E.}~\bibnamefont {Komatsu}} \emph {et~al.} (\bibinfo {collaboration} {WMAP}),\ }\bibfield  {title} {\bibinfo {title} {{Seven-Year Wilkinson Microwave Anisotropy Probe (WMAP) Observations: Cosmological Interpretation}},\ }\href {https://doi.org/10.1088/0067-0049/192/2/18} {\bibfield  {journal} {\bibinfo  {journal} {Astrophys. J. Suppl.}\ }\textbf {\bibinfo {volume} {192}},\ \bibinfo {pages} {18} (\bibinfo {year} {2011})},\ \Eprint {https://arxiv.org/abs/1001.4538} {arXiv:1001.4538 [astro-ph.CO]} \BibitemShut {NoStop}%
\bibitem [{\citenamefont {Akrami}\ \emph {et~al.}(2020)\citenamefont {Akrami} \emph {et~al.}}]{Planck:2018jri}%
  \BibitemOpen
  \bibfield  {author} {\bibinfo {author} {\bibfnamefont {Y.}~\bibnamefont {Akrami}} \emph {et~al.} (\bibinfo {collaboration} {Planck}),\ }\bibfield  {title} {\bibinfo {title} {{Planck 2018 results. X. Constraints on inflation}},\ }\href {https://doi.org/10.1051/0004-6361/201833887} {\bibfield  {journal} {\bibinfo  {journal} {Astron. Astrophys.}\ }\textbf {\bibinfo {volume} {641}},\ \bibinfo {pages} {A10} (\bibinfo {year} {2020})},\ \Eprint {https://arxiv.org/abs/1807.06211} {arXiv:1807.06211 [astro-ph.CO]} \BibitemShut {NoStop}%
\bibitem [{\citenamefont {Bennett}\ \emph {et~al.}(2013)\citenamefont {Bennett} \emph {et~al.}}]{WMAP:2012fli}%
  \BibitemOpen
  \bibfield  {author} {\bibinfo {author} {\bibfnamefont {C.~L.}\ \bibnamefont {Bennett}} \emph {et~al.} (\bibinfo {collaboration} {WMAP}),\ }\bibfield  {title} {\bibinfo {title} {{Nine-Year Wilkinson Microwave Anisotropy Probe (WMAP) Observations: Final Maps and Results}},\ }\href {https://doi.org/10.1088/0067-0049/208/2/20} {\bibfield  {journal} {\bibinfo  {journal} {Astrophys. J. Suppl.}\ }\textbf {\bibinfo {volume} {208}},\ \bibinfo {pages} {20} (\bibinfo {year} {2013})},\ \Eprint {https://arxiv.org/abs/1212.5225} {arXiv:1212.5225 [astro-ph.CO]} \BibitemShut {NoStop}%
\bibitem [{\citenamefont {Frieman}\ \emph {et~al.}(2008)\citenamefont {Frieman}, \citenamefont {Turner},\ and\ \citenamefont {Huterer}}]{Frieman:2008sn}%
  \BibitemOpen
  \bibfield  {author} {\bibinfo {author} {\bibfnamefont {J.}~\bibnamefont {Frieman}}, \bibinfo {author} {\bibfnamefont {M.}~\bibnamefont {Turner}},\ and\ \bibinfo {author} {\bibfnamefont {D.}~\bibnamefont {Huterer}},\ }\bibfield  {title} {\bibinfo {title} {{Dark Energy and the Accelerating Universe}},\ }\href {https://doi.org/10.1146/annurev.astro.46.060407.145243} {\bibfield  {journal} {\bibinfo  {journal} {Ann. Rev. Astron. Astrophys.}\ }\textbf {\bibinfo {volume} {46}},\ \bibinfo {pages} {385} (\bibinfo {year} {2008})},\ \Eprint {https://arxiv.org/abs/0803.0982} {arXiv:0803.0982 [astro-ph]} \BibitemShut {NoStop}%
\bibitem [{\citenamefont {Riess}\ \emph {et~al.}(1998)\citenamefont {Riess} \emph {et~al.}}]{SupernovaSearchTeam:1998fmf}%
  \BibitemOpen
  \bibfield  {author} {\bibinfo {author} {\bibfnamefont {A.~G.}\ \bibnamefont {Riess}} \emph {et~al.} (\bibinfo {collaboration} {Supernova Search Team}),\ }\bibfield  {title} {\bibinfo {title} {{Observational evidence from supernovae for an accelerating universe and a cosmological constant}},\ }\href {https://doi.org/10.1086/300499} {\bibfield  {journal} {\bibinfo  {journal} {Astron. J.}\ }\textbf {\bibinfo {volume} {116}},\ \bibinfo {pages} {1009} (\bibinfo {year} {1998})},\ \Eprint {https://arxiv.org/abs/astro-ph/9805201} {arXiv:astro-ph/9805201} \BibitemShut {NoStop}%
\bibitem [{\citenamefont {Perlmutter}\ \emph {et~al.}(1999)\citenamefont {Perlmutter} \emph {et~al.}}]{SupernovaCosmologyProject:1998vns}%
  \BibitemOpen
  \bibfield  {author} {\bibinfo {author} {\bibfnamefont {S.}~\bibnamefont {Perlmutter}} \emph {et~al.} (\bibinfo {collaboration} {Supernova Cosmology Project}),\ }\bibfield  {title} {\bibinfo {title} {{Measurements of $\Omega$ and $\Lambda$ from 42 high redshift supernovae}},\ }\href {https://doi.org/10.1086/307221} {\bibfield  {journal} {\bibinfo  {journal} {Astrophys. J.}\ }\textbf {\bibinfo {volume} {517}},\ \bibinfo {pages} {565} (\bibinfo {year} {1999})},\ \Eprint {https://arxiv.org/abs/astro-ph/9812133} {arXiv:astro-ph/9812133} \BibitemShut {NoStop}%
\bibitem [{\citenamefont {Nojiri}\ \emph {et~al.}(2017)\citenamefont {Nojiri}, \citenamefont {Odintsov},\ and\ \citenamefont {Oikonomou}}]{Nojiri:2017ncd}%
  \BibitemOpen
  \bibfield  {author} {\bibinfo {author} {\bibfnamefont {S.}~\bibnamefont {Nojiri}}, \bibinfo {author} {\bibfnamefont {S.~D.}\ \bibnamefont {Odintsov}},\ and\ \bibinfo {author} {\bibfnamefont {V.~K.}\ \bibnamefont {Oikonomou}},\ }\bibfield  {title} {\bibinfo {title} {{Modified Gravity Theories on a Nutshell: Inflation, Bounce and Late-time Evolution}},\ }\href {https://doi.org/10.1016/j.physrep.2017.06.001} {\bibfield  {journal} {\bibinfo  {journal} {Phys. Rept.}\ }\textbf {\bibinfo {volume} {692}},\ \bibinfo {pages} {1} (\bibinfo {year} {2017})},\ \Eprint {https://arxiv.org/abs/1705.11098} {arXiv:1705.11098 [gr-qc]} \BibitemShut {NoStop}%
\bibitem [{\citenamefont {Suzuki}\ \emph {et~al.}(2012)\citenamefont {Suzuki} \emph {et~al.}}]{SupernovaCosmologyProject:2011ycw}%
  \BibitemOpen
  \bibfield  {author} {\bibinfo {author} {\bibfnamefont {N.}~\bibnamefont {Suzuki}} \emph {et~al.} (\bibinfo {collaboration} {Supernova Cosmology Project}),\ }\bibfield  {title} {\bibinfo {title} {{The Hubble Space Telescope Cluster Supernova Survey: V. Improving the Dark Energy Constraints Above z\ensuremath{>}1 and Building an Early-Type-Hosted Supernova Sample}},\ }\href {https://doi.org/10.1088/0004-637X/746/1/85} {\bibfield  {journal} {\bibinfo  {journal} {Astrophys. J.}\ }\textbf {\bibinfo {volume} {746}},\ \bibinfo {pages} {85} (\bibinfo {year} {2012})},\ \Eprint {https://arxiv.org/abs/1105.3470} {arXiv:1105.3470 [astro-ph.CO]} \BibitemShut {NoStop}%
\bibitem [{\citenamefont {Kowalski}\ \emph {et~al.}(2008)\citenamefont {Kowalski} \emph {et~al.}}]{SupernovaCosmologyProject:2008ojh}%
  \BibitemOpen
  \bibfield  {author} {\bibinfo {author} {\bibfnamefont {M.}~\bibnamefont {Kowalski}} \emph {et~al.} (\bibinfo {collaboration} {Supernova Cosmology Project}),\ }\bibfield  {title} {\bibinfo {title} {{Improved Cosmological Constraints from New, Old and Combined Supernova Datasets}},\ }\href {https://doi.org/10.1086/589937} {\bibfield  {journal} {\bibinfo  {journal} {Astrophys. J.}\ }\textbf {\bibinfo {volume} {686}},\ \bibinfo {pages} {749} (\bibinfo {year} {2008})},\ \Eprint {https://arxiv.org/abs/0804.4142} {arXiv:0804.4142 [astro-ph]} \BibitemShut {NoStop}%
\bibitem [{\citenamefont {Aghanim}\ \emph {et~al.}(2020{\natexlab{b}})\citenamefont {Aghanim} \emph {et~al.}}]{Planck:2018nkj}%
  \BibitemOpen
  \bibfield  {author} {\bibinfo {author} {\bibfnamefont {N.}~\bibnamefont {Aghanim}} \emph {et~al.} (\bibinfo {collaboration} {Planck}),\ }\bibfield  {title} {\bibinfo {title} {{Planck 2018 results. I. Overview and the cosmological legacy of Planck}},\ }\href {https://doi.org/10.1051/0004-6361/201833880} {\bibfield  {journal} {\bibinfo  {journal} {Astron. Astrophys.}\ }\textbf {\bibinfo {volume} {641}},\ \bibinfo {pages} {A1} (\bibinfo {year} {2020}{\natexlab{b}})},\ \Eprint {https://arxiv.org/abs/1807.06205} {arXiv:1807.06205 [astro-ph.CO]} \BibitemShut {NoStop}%
\bibitem [{\citenamefont {Feng}\ \emph {et~al.}(2005)\citenamefont {Feng}, \citenamefont {Wang},\ and\ \citenamefont {Zhang}}]{Feng:2004ad}%
  \BibitemOpen
  \bibfield  {author} {\bibinfo {author} {\bibfnamefont {B.}~\bibnamefont {Feng}}, \bibinfo {author} {\bibfnamefont {X.-L.}\ \bibnamefont {Wang}},\ and\ \bibinfo {author} {\bibfnamefont {X.-M.}\ \bibnamefont {Zhang}},\ }\bibfield  {title} {\bibinfo {title} {{Dark energy constraints from the cosmic age and supernova}},\ }\href {https://doi.org/10.1016/j.physletb.2004.12.071} {\bibfield  {journal} {\bibinfo  {journal} {Phys. Lett. B}\ }\textbf {\bibinfo {volume} {607}},\ \bibinfo {pages} {35} (\bibinfo {year} {2005})},\ \Eprint {https://arxiv.org/abs/astro-ph/0404224} {arXiv:astro-ph/0404224} \BibitemShut {NoStop}%
\bibitem [{\citenamefont {Weinberg}\ \emph {et~al.}(2013)\citenamefont {Weinberg}, \citenamefont {Mortonson}, \citenamefont {Eisenstein}, \citenamefont {Hirata}, \citenamefont {Riess},\ and\ \citenamefont {Rozo}}]{Weinberg:2013agg}%
  \BibitemOpen
  \bibfield  {author} {\bibinfo {author} {\bibfnamefont {D.~H.}\ \bibnamefont {Weinberg}}, \bibinfo {author} {\bibfnamefont {M.~J.}\ \bibnamefont {Mortonson}}, \bibinfo {author} {\bibfnamefont {D.~J.}\ \bibnamefont {Eisenstein}}, \bibinfo {author} {\bibfnamefont {C.}~\bibnamefont {Hirata}}, \bibinfo {author} {\bibfnamefont {A.~G.}\ \bibnamefont {Riess}},\ and\ \bibinfo {author} {\bibfnamefont {E.}~\bibnamefont {Rozo}},\ }\bibfield  {title} {\bibinfo {title} {{Observational Probes of Cosmic Acceleration}},\ }\href {https://doi.org/10.1016/j.physrep.2013.05.001} {\bibfield  {journal} {\bibinfo  {journal} {Phys. Rept.}\ }\textbf {\bibinfo {volume} {530}},\ \bibinfo {pages} {87} (\bibinfo {year} {2013})},\ \Eprint {https://arxiv.org/abs/1201.2434} {arXiv:1201.2434 [astro-ph.CO]} \BibitemShut {NoStop}%
\bibitem [{\citenamefont {Seljak}\ \emph {et~al.}(2005)\citenamefont {Seljak} \emph {et~al.}}]{SDSS:2004kqt}%
  \BibitemOpen
  \bibfield  {author} {\bibinfo {author} {\bibfnamefont {U.}~\bibnamefont {Seljak}} \emph {et~al.} (\bibinfo {collaboration} {SDSS}),\ }\bibfield  {title} {\bibinfo {title} {{Cosmological parameter analysis including SDSS Ly-alpha forest and galaxy bias: Constraints on the primordial spectrum of fluctuations, neutrino mass, and dark energy}},\ }\href {https://doi.org/10.1103/PhysRevD.71.103515} {\bibfield  {journal} {\bibinfo  {journal} {Phys. Rev. D}\ }\textbf {\bibinfo {volume} {71}},\ \bibinfo {pages} {103515} (\bibinfo {year} {2005})},\ \Eprint {https://arxiv.org/abs/astro-ph/0407372} {arXiv:astro-ph/0407372} \BibitemShut {NoStop}%
\bibitem [{\citenamefont {Allen}\ \emph {et~al.}(2011)\citenamefont {Allen}, \citenamefont {Evrard},\ and\ \citenamefont {Mantz}}]{Allen:2011zs}%
  \BibitemOpen
  \bibfield  {author} {\bibinfo {author} {\bibfnamefont {S.~W.}\ \bibnamefont {Allen}}, \bibinfo {author} {\bibfnamefont {A.~E.}\ \bibnamefont {Evrard}},\ and\ \bibinfo {author} {\bibfnamefont {A.~B.}\ \bibnamefont {Mantz}},\ }\bibfield  {title} {\bibinfo {title} {{Cosmological Parameters from Observations of Galaxy Clusters}},\ }\href {https://doi.org/10.1146/annurev-astro-081710-102514} {\bibfield  {journal} {\bibinfo  {journal} {Ann. Rev. Astron. Astrophys.}\ }\textbf {\bibinfo {volume} {49}},\ \bibinfo {pages} {409} (\bibinfo {year} {2011})},\ \Eprint {https://arxiv.org/abs/1103.4829} {arXiv:1103.4829 [astro-ph.CO]} \BibitemShut {NoStop}%
\bibitem [{\citenamefont {Copeland}\ \emph {et~al.}(2006)\citenamefont {Copeland}, \citenamefont {Sami},\ and\ \citenamefont {Tsujikawa}}]{Copeland:2006wr}%
  \BibitemOpen
  \bibfield  {author} {\bibinfo {author} {\bibfnamefont {E.~J.}\ \bibnamefont {Copeland}}, \bibinfo {author} {\bibfnamefont {M.}~\bibnamefont {Sami}},\ and\ \bibinfo {author} {\bibfnamefont {S.}~\bibnamefont {Tsujikawa}},\ }\bibfield  {title} {\bibinfo {title} {{Dynamics of dark energy}},\ }\href {https://doi.org/10.1142/S021827180600942X} {\bibfield  {journal} {\bibinfo  {journal} {Int. J. Mod. Phys. D}\ }\textbf {\bibinfo {volume} {15}},\ \bibinfo {pages} {1753} (\bibinfo {year} {2006})},\ \Eprint {https://arxiv.org/abs/hep-th/0603057} {arXiv:hep-th/0603057} \BibitemShut {NoStop}%
\bibitem [{\citenamefont {Joyce}\ \emph {et~al.}(2015)\citenamefont {Joyce}, \citenamefont {Jain}, \citenamefont {Khoury},\ and\ \citenamefont {Trodden}}]{Joyce:2014kja}%
  \BibitemOpen
  \bibfield  {author} {\bibinfo {author} {\bibfnamefont {A.}~\bibnamefont {Joyce}}, \bibinfo {author} {\bibfnamefont {B.}~\bibnamefont {Jain}}, \bibinfo {author} {\bibfnamefont {J.}~\bibnamefont {Khoury}},\ and\ \bibinfo {author} {\bibfnamefont {M.}~\bibnamefont {Trodden}},\ }\bibfield  {title} {\bibinfo {title} {{Beyond the Cosmological Standard Model}},\ }\href {https://doi.org/10.1016/j.physrep.2014.12.002} {\bibfield  {journal} {\bibinfo  {journal} {Phys. Rept.}\ }\textbf {\bibinfo {volume} {568}},\ \bibinfo {pages} {1} (\bibinfo {year} {2015})},\ \Eprint {https://arxiv.org/abs/1407.0059} {arXiv:1407.0059 [astro-ph.CO]} \BibitemShut {NoStop}%
\bibitem [{\citenamefont {Peebles}\ and\ \citenamefont {Ratra}(2003)}]{Peebles:2002gy}%
  \BibitemOpen
  \bibfield  {author} {\bibinfo {author} {\bibfnamefont {P.~J.~E.}\ \bibnamefont {Peebles}}\ and\ \bibinfo {author} {\bibfnamefont {B.}~\bibnamefont {Ratra}},\ }\bibfield  {title} {\bibinfo {title} {{The Cosmological Constant and Dark Energy}},\ }\href {https://doi.org/10.1103/RevModPhys.75.559} {\bibfield  {journal} {\bibinfo  {journal} {Rev. Mod. Phys.}\ }\textbf {\bibinfo {volume} {75}},\ \bibinfo {pages} {559} (\bibinfo {year} {2003})},\ \Eprint {https://arxiv.org/abs/astro-ph/0207347} {arXiv:astro-ph/0207347} \BibitemShut {NoStop}%
\bibitem [{\citenamefont {Padmanabhan}(2003)}]{Padmanabhan:2002ji}%
  \BibitemOpen
  \bibfield  {author} {\bibinfo {author} {\bibfnamefont {T.}~\bibnamefont {Padmanabhan}},\ }\bibfield  {title} {\bibinfo {title} {{Cosmological constant: The Weight of the vacuum}},\ }\href {https://doi.org/10.1016/S0370-1573(03)00120-0} {\bibfield  {journal} {\bibinfo  {journal} {Phys. Rept.}\ }\textbf {\bibinfo {volume} {380}},\ \bibinfo {pages} {235} (\bibinfo {year} {2003})},\ \Eprint {https://arxiv.org/abs/hep-th/0212290} {arXiv:hep-th/0212290} \BibitemShut {NoStop}%
\bibitem [{\citenamefont {Boylan-Kolchin}\ \emph {et~al.}(2012)\citenamefont {Boylan-Kolchin}, \citenamefont {Bullock},\ and\ \citenamefont {Kaplinghat}}]{Boylan-Kolchin:2011lmk}%
  \BibitemOpen
  \bibfield  {author} {\bibinfo {author} {\bibfnamefont {M.}~\bibnamefont {Boylan-Kolchin}}, \bibinfo {author} {\bibfnamefont {J.~S.}\ \bibnamefont {Bullock}},\ and\ \bibinfo {author} {\bibfnamefont {M.}~\bibnamefont {Kaplinghat}},\ }\bibfield  {title} {\bibinfo {title} {{The Milky Way's bright satellites as an apparent failure of LCDM}},\ }\href {https://doi.org/10.1111/j.1365-2966.2012.20695.x} {\bibfield  {journal} {\bibinfo  {journal} {Mon. Not. Roy. Astron. Soc.}\ }\textbf {\bibinfo {volume} {422}},\ \bibinfo {pages} {1203} (\bibinfo {year} {2012})},\ \Eprint {https://arxiv.org/abs/1111.2048} {arXiv:1111.2048 [astro-ph.CO]} \BibitemShut {NoStop}%
\bibitem [{\citenamefont {Perivolaropoulos}\ and\ \citenamefont {Skara}(2022)}]{Perivolaropoulos:2021jda}%
  \BibitemOpen
  \bibfield  {author} {\bibinfo {author} {\bibfnamefont {L.}~\bibnamefont {Perivolaropoulos}}\ and\ \bibinfo {author} {\bibfnamefont {F.}~\bibnamefont {Skara}},\ }\bibfield  {title} {\bibinfo {title} {{Challenges for \ensuremath{\Lambda}CDM: An update}},\ }\href {https://doi.org/10.1016/j.newar.2022.101659} {\bibfield  {journal} {\bibinfo  {journal} {New Astron. Rev.}\ }\textbf {\bibinfo {volume} {95}},\ \bibinfo {pages} {101659} (\bibinfo {year} {2022})},\ \Eprint {https://arxiv.org/abs/2105.05208} {arXiv:2105.05208 [astro-ph.CO]} \BibitemShut {NoStop}%
\bibitem [{\citenamefont {Vagnozzi}(2023)}]{Vagnozzi:2023nrq}%
  \BibitemOpen
  \bibfield  {author} {\bibinfo {author} {\bibfnamefont {S.}~\bibnamefont {Vagnozzi}},\ }\bibfield  {title} {\bibinfo {title} {{Seven Hints That Early-Time New Physics Alone Is Not Sufficient to Solve the Hubble Tension}},\ }\href {https://doi.org/10.3390/universe9090393} {\bibfield  {journal} {\bibinfo  {journal} {Universe}\ }\textbf {\bibinfo {volume} {9}},\ \bibinfo {pages} {393} (\bibinfo {year} {2023})},\ \Eprint {https://arxiv.org/abs/2308.16628} {arXiv:2308.16628 [astro-ph.CO]} \BibitemShut {NoStop}%
\bibitem [{\citenamefont {L\'opez-Corredoira}(2017)}]{Lopez-Corredoira:2017rqn}%
  \BibitemOpen
  \bibfield  {author} {\bibinfo {author} {\bibfnamefont {M.}~\bibnamefont {L\'opez-Corredoira}},\ }\bibfield  {title} {\bibinfo {title} {{Tests and problems of the standard model in Cosmology}},\ }\href {https://doi.org/10.1007/s10701-017-0073-8} {\bibfield  {journal} {\bibinfo  {journal} {Found. Phys.}\ }\textbf {\bibinfo {volume} {47}},\ \bibinfo {pages} {711} (\bibinfo {year} {2017})},\ \Eprint {https://arxiv.org/abs/1701.08720} {arXiv:1701.08720 [astro-ph.CO]} \BibitemShut {NoStop}%
\bibitem [{\citenamefont {Past\'en}\ and\ \citenamefont {C\'ardenas}(2023)}]{Pasten:2023rpc}%
  \BibitemOpen
  \bibfield  {author} {\bibinfo {author} {\bibfnamefont {E.}~\bibnamefont {Past\'en}}\ and\ \bibinfo {author} {\bibfnamefont {V.~H.}\ \bibnamefont {C\'ardenas}},\ }\bibfield  {title} {\bibinfo {title} {{Testing \ensuremath{\Lambda}CDM cosmology in a binned universe: Anomalies in the deceleration parameter}},\ }\href {https://doi.org/10.1016/j.dark.2023.101224} {\bibfield  {journal} {\bibinfo  {journal} {Phys. Dark Univ.}\ }\textbf {\bibinfo {volume} {40}},\ \bibinfo {pages} {101224} (\bibinfo {year} {2023})},\ \Eprint {https://arxiv.org/abs/2301.10740} {arXiv:2301.10740 [astro-ph.CO]} \BibitemShut {NoStop}%
\bibitem [{\citenamefont {Wang}(2022)}]{Wang:2022xdw}%
  \BibitemOpen
  \bibfield  {author} {\bibinfo {author} {\bibfnamefont {D.}~\bibnamefont {Wang}},\ }\bibfield  {title} {\bibinfo {title} {{Pantheon+ constraints on dark energy and modified gravity: An evidence of dynamical dark energy}},\ }\href {https://doi.org/10.1103/PhysRevD.106.063515} {\bibfield  {journal} {\bibinfo  {journal} {Phys. Rev. D}\ }\textbf {\bibinfo {volume} {106}},\ \bibinfo {pages} {063515} (\bibinfo {year} {2022})},\ \Eprint {https://arxiv.org/abs/2207.07164} {arXiv:2207.07164 [astro-ph.CO]} \BibitemShut {NoStop}%
\bibitem [{\citenamefont {Zhao}\ \emph {et~al.}(2017)\citenamefont {Zhao} \emph {et~al.}}]{Zhao:2017cud}%
  \BibitemOpen
  \bibfield  {author} {\bibinfo {author} {\bibfnamefont {G.-B.}\ \bibnamefont {Zhao}} \emph {et~al.},\ }\bibfield  {title} {\bibinfo {title} {{Dynamical dark energy in light of the latest observations}},\ }\href {https://doi.org/10.1038/s41550-017-0216-z} {\bibfield  {journal} {\bibinfo  {journal} {Nature Astron.}\ }\textbf {\bibinfo {volume} {1}},\ \bibinfo {pages} {627} (\bibinfo {year} {2017})},\ \Eprint {https://arxiv.org/abs/1701.08165} {arXiv:1701.08165 [astro-ph.CO]} \BibitemShut {NoStop}%
\bibitem [{\citenamefont {Upadhye}\ \emph {et~al.}(2005)\citenamefont {Upadhye}, \citenamefont {Ishak},\ and\ \citenamefont {Steinhardt}}]{Upadhye:2004hh}%
  \BibitemOpen
  \bibfield  {author} {\bibinfo {author} {\bibfnamefont {A.}~\bibnamefont {Upadhye}}, \bibinfo {author} {\bibfnamefont {M.}~\bibnamefont {Ishak}},\ and\ \bibinfo {author} {\bibfnamefont {P.~J.}\ \bibnamefont {Steinhardt}},\ }\bibfield  {title} {\bibinfo {title} {{Dynamical dark energy: Current constraints and forecasts}},\ }\href {https://doi.org/10.1103/PhysRevD.72.063501} {\bibfield  {journal} {\bibinfo  {journal} {Phys. Rev. D}\ }\textbf {\bibinfo {volume} {72}},\ \bibinfo {pages} {063501} (\bibinfo {year} {2005})},\ \Eprint {https://arxiv.org/abs/astro-ph/0411803} {arXiv:astro-ph/0411803} \BibitemShut {NoStop}%
\bibitem [{\citenamefont {Di~Valentino}\ \emph {et~al.}(2017)\citenamefont {Di~Valentino}, \citenamefont {Melchiorri}, \citenamefont {Linder},\ and\ \citenamefont {Silk}}]{DiValentino:2017zyq}%
  \BibitemOpen
  \bibfield  {author} {\bibinfo {author} {\bibfnamefont {E.}~\bibnamefont {Di~Valentino}}, \bibinfo {author} {\bibfnamefont {A.}~\bibnamefont {Melchiorri}}, \bibinfo {author} {\bibfnamefont {E.~V.}\ \bibnamefont {Linder}},\ and\ \bibinfo {author} {\bibfnamefont {J.}~\bibnamefont {Silk}},\ }\bibfield  {title} {\bibinfo {title} {{Constraining Dark Energy Dynamics in Extended Parameter Space}},\ }\href {https://doi.org/10.1103/PhysRevD.96.023523} {\bibfield  {journal} {\bibinfo  {journal} {Phys. Rev. D}\ }\textbf {\bibinfo {volume} {96}},\ \bibinfo {pages} {023523} (\bibinfo {year} {2017})},\ \Eprint {https://arxiv.org/abs/1704.00762} {arXiv:1704.00762 [astro-ph.CO]} \BibitemShut {NoStop}%
\bibitem [{\citenamefont {Capozziello}\ and\ \citenamefont {De~Laurentis}(2011)}]{Capozziello:2011et}%
  \BibitemOpen
  \bibfield  {author} {\bibinfo {author} {\bibfnamefont {S.}~\bibnamefont {Capozziello}}\ and\ \bibinfo {author} {\bibfnamefont {M.}~\bibnamefont {De~Laurentis}},\ }\bibfield  {title} {\bibinfo {title} {{Extended Theories of Gravity}},\ }\href {https://doi.org/10.1016/j.physrep.2011.09.003} {\bibfield  {journal} {\bibinfo  {journal} {Phys. Rept.}\ }\textbf {\bibinfo {volume} {509}},\ \bibinfo {pages} {167} (\bibinfo {year} {2011})},\ \Eprint {https://arxiv.org/abs/1108.6266} {arXiv:1108.6266 [gr-qc]} \BibitemShut {NoStop}%
\bibitem [{\citenamefont {Nojiri}\ and\ \citenamefont {Odintsov}(2011)}]{Nojiri:2010wj}%
  \BibitemOpen
  \bibfield  {author} {\bibinfo {author} {\bibfnamefont {S.}~\bibnamefont {Nojiri}}\ and\ \bibinfo {author} {\bibfnamefont {S.~D.}\ \bibnamefont {Odintsov}},\ }\bibfield  {title} {\bibinfo {title} {{Unified cosmic history in modified gravity: from F(R) theory to Lorentz non-invariant models}},\ }\href {https://doi.org/10.1016/j.physrep.2011.04.001} {\bibfield  {journal} {\bibinfo  {journal} {Phys. Rept.}\ }\textbf {\bibinfo {volume} {505}},\ \bibinfo {pages} {59} (\bibinfo {year} {2011})},\ \Eprint {https://arxiv.org/abs/1011.0544} {arXiv:1011.0544 [gr-qc]} \BibitemShut {NoStop}%
\bibitem [{\citenamefont {Clifton}\ \emph {et~al.}(2012)\citenamefont {Clifton}, \citenamefont {Ferreira}, \citenamefont {Padilla},\ and\ \citenamefont {Skordis}}]{Clifton:2011jh}%
  \BibitemOpen
  \bibfield  {author} {\bibinfo {author} {\bibfnamefont {T.}~\bibnamefont {Clifton}}, \bibinfo {author} {\bibfnamefont {P.~G.}\ \bibnamefont {Ferreira}}, \bibinfo {author} {\bibfnamefont {A.}~\bibnamefont {Padilla}},\ and\ \bibinfo {author} {\bibfnamefont {C.}~\bibnamefont {Skordis}},\ }\bibfield  {title} {\bibinfo {title} {{Modified Gravity and Cosmology}},\ }\href {https://doi.org/10.1016/j.physrep.2012.01.001} {\bibfield  {journal} {\bibinfo  {journal} {Phys. Rept.}\ }\textbf {\bibinfo {volume} {513}},\ \bibinfo {pages} {1} (\bibinfo {year} {2012})},\ \Eprint {https://arxiv.org/abs/1106.2476} {arXiv:1106.2476 [astro-ph.CO]} \BibitemShut {NoStop}%
\bibitem [{\citenamefont {Nojiri}\ and\ \citenamefont {Odintsov}(2006{\natexlab{a}})}]{Nojiri:2006ri}%
  \BibitemOpen
  \bibfield  {author} {\bibinfo {author} {\bibfnamefont {S.}~\bibnamefont {Nojiri}}\ and\ \bibinfo {author} {\bibfnamefont {S.~D.}\ \bibnamefont {Odintsov}},\ }\bibfield  {title} {\bibinfo {title} {{Introduction to modified gravity and gravitational alternative for dark energy}},\ }\href {https://doi.org/10.1142/S0219887807001928} {\bibfield  {journal} {\bibinfo  {journal} {eConf}\ }\textbf {\bibinfo {volume} {C0602061}},\ \bibinfo {pages} {06} (\bibinfo {year} {2006}{\natexlab{a}})},\ \Eprint {https://arxiv.org/abs/hep-th/0601213} {arXiv:hep-th/0601213} \BibitemShut {NoStop}%
\bibitem [{\citenamefont {Tsujikawa}(2010)}]{Tsujikawa:2010zza}%
  \BibitemOpen
  \bibfield  {author} {\bibinfo {author} {\bibfnamefont {S.}~\bibnamefont {Tsujikawa}},\ }\bibfield  {title} {\bibinfo {title} {{Modified gravity models of dark energy}},\ }\href {https://doi.org/10.1007/978-3-642-10598-2_3} {\bibfield  {journal} {\bibinfo  {journal} {Lect. Notes Phys.}\ }\textbf {\bibinfo {volume} {800}},\ \bibinfo {pages} {99} (\bibinfo {year} {2010})},\ \Eprint {https://arxiv.org/abs/1101.0191} {arXiv:1101.0191 [gr-qc]} \BibitemShut {NoStop}%
\bibitem [{\citenamefont {De~Felice}\ and\ \citenamefont {Tsujikawa}(2010)}]{DeFelice:2010aj}%
  \BibitemOpen
  \bibfield  {author} {\bibinfo {author} {\bibfnamefont {A.}~\bibnamefont {De~Felice}}\ and\ \bibinfo {author} {\bibfnamefont {S.}~\bibnamefont {Tsujikawa}},\ }\bibfield  {title} {\bibinfo {title} {{f(R) theories}},\ }\href {https://doi.org/10.12942/lrr-2010-3} {\bibfield  {journal} {\bibinfo  {journal} {Living Rev. Rel.}\ }\textbf {\bibinfo {volume} {13}},\ \bibinfo {pages} {3} (\bibinfo {year} {2010})},\ \Eprint {https://arxiv.org/abs/1002.4928} {arXiv:1002.4928 [gr-qc]} \BibitemShut {NoStop}%
\bibitem [{\citenamefont {Amendola}\ \emph {et~al.}(2007)\citenamefont {Amendola}, \citenamefont {Polarski},\ and\ \citenamefont {Tsujikawa}}]{Amendola:2006kh}%
  \BibitemOpen
  \bibfield  {author} {\bibinfo {author} {\bibfnamefont {L.}~\bibnamefont {Amendola}}, \bibinfo {author} {\bibfnamefont {D.}~\bibnamefont {Polarski}},\ and\ \bibinfo {author} {\bibfnamefont {S.}~\bibnamefont {Tsujikawa}},\ }\bibfield  {title} {\bibinfo {title} {{Are f(R) dark energy models cosmologically viable ?}},\ }\href {https://doi.org/10.1103/PhysRevLett.98.131302} {\bibfield  {journal} {\bibinfo  {journal} {Phys. Rev. Lett.}\ }\textbf {\bibinfo {volume} {98}},\ \bibinfo {pages} {131302} (\bibinfo {year} {2007})},\ \Eprint {https://arxiv.org/abs/astro-ph/0603703} {arXiv:astro-ph/0603703} \BibitemShut {NoStop}%
\bibitem [{\citenamefont {Amarzguioui}\ \emph {et~al.}(2006)\citenamefont {Amarzguioui}, \citenamefont {Elgaroy}, \citenamefont {Mota},\ and\ \citenamefont {Multamaki}}]{Amarzguioui:2005zq}%
  \BibitemOpen
  \bibfield  {author} {\bibinfo {author} {\bibfnamefont {M.}~\bibnamefont {Amarzguioui}}, \bibinfo {author} {\bibfnamefont {O.}~\bibnamefont {Elgaroy}}, \bibinfo {author} {\bibfnamefont {D.~F.}\ \bibnamefont {Mota}},\ and\ \bibinfo {author} {\bibfnamefont {T.}~\bibnamefont {Multamaki}},\ }\bibfield  {title} {\bibinfo {title} {{Cosmological constraints on f(r) gravity theories within the palatini approach}},\ }\href {https://doi.org/10.1051/0004-6361:20064994} {\bibfield  {journal} {\bibinfo  {journal} {Astron. Astrophys.}\ }\textbf {\bibinfo {volume} {454}},\ \bibinfo {pages} {707} (\bibinfo {year} {2006})},\ \Eprint {https://arxiv.org/abs/astro-ph/0510519} {arXiv:astro-ph/0510519} \BibitemShut {NoStop}%
\bibitem [{\citenamefont {Sotiriou}\ and\ \citenamefont {Faraoni}(2010)}]{Sotiriou:2008rp}%
  \BibitemOpen
  \bibfield  {author} {\bibinfo {author} {\bibfnamefont {T.~P.}\ \bibnamefont {Sotiriou}}\ and\ \bibinfo {author} {\bibfnamefont {V.}~\bibnamefont {Faraoni}},\ }\bibfield  {title} {\bibinfo {title} {{f(R) Theories Of Gravity}},\ }\href {https://doi.org/10.1103/RevModPhys.82.451} {\bibfield  {journal} {\bibinfo  {journal} {Rev. Mod. Phys.}\ }\textbf {\bibinfo {volume} {82}},\ \bibinfo {pages} {451} (\bibinfo {year} {2010})},\ \Eprint {https://arxiv.org/abs/0805.1726} {arXiv:0805.1726 [gr-qc]} \BibitemShut {NoStop}%
\bibitem [{\citenamefont {Nojiri}\ and\ \citenamefont {Odintsov}(2006{\natexlab{b}})}]{Nojiri:2006gh}%
  \BibitemOpen
  \bibfield  {author} {\bibinfo {author} {\bibfnamefont {S.}~\bibnamefont {Nojiri}}\ and\ \bibinfo {author} {\bibfnamefont {S.~D.}\ \bibnamefont {Odintsov}},\ }\bibfield  {title} {\bibinfo {title} {{Modified f(R) gravity consistent with realistic cosmology: From matter dominated epoch to dark energy universe}},\ }\href {https://doi.org/10.1103/PhysRevD.74.086005} {\bibfield  {journal} {\bibinfo  {journal} {Phys. Rev. D}\ }\textbf {\bibinfo {volume} {74}},\ \bibinfo {pages} {086005} (\bibinfo {year} {2006}{\natexlab{b}})},\ \Eprint {https://arxiv.org/abs/hep-th/0608008} {arXiv:hep-th/0608008} \BibitemShut {NoStop}%
\bibitem [{\citenamefont {Harko}\ \emph {et~al.}(2011)\citenamefont {Harko}, \citenamefont {Lobo}, \citenamefont {Nojiri},\ and\ \citenamefont {Odintsov}}]{Harko:2011kv}%
  \BibitemOpen
  \bibfield  {author} {\bibinfo {author} {\bibfnamefont {T.}~\bibnamefont {Harko}}, \bibinfo {author} {\bibfnamefont {F.~S.~N.}\ \bibnamefont {Lobo}}, \bibinfo {author} {\bibfnamefont {S.}~\bibnamefont {Nojiri}},\ and\ \bibinfo {author} {\bibfnamefont {S.~D.}\ \bibnamefont {Odintsov}},\ }\bibfield  {title} {\bibinfo {title} {{$f(R,T)$ gravity}},\ }\href {https://doi.org/10.1103/PhysRevD.84.024020} {\bibfield  {journal} {\bibinfo  {journal} {Phys. Rev. D}\ }\textbf {\bibinfo {volume} {84}},\ \bibinfo {pages} {024020} (\bibinfo {year} {2011})},\ \Eprint {https://arxiv.org/abs/1104.2669} {arXiv:1104.2669 [gr-qc]} \BibitemShut {NoStop}%
\bibitem [{\citenamefont {Cai}\ \emph {et~al.}(2016)\citenamefont {Cai}, \citenamefont {Capozziello}, \citenamefont {De~Laurentis},\ and\ \citenamefont {Saridakis}}]{Cai:2015emx}%
  \BibitemOpen
  \bibfield  {author} {\bibinfo {author} {\bibfnamefont {Y.-F.}\ \bibnamefont {Cai}}, \bibinfo {author} {\bibfnamefont {S.}~\bibnamefont {Capozziello}}, \bibinfo {author} {\bibfnamefont {M.}~\bibnamefont {De~Laurentis}},\ and\ \bibinfo {author} {\bibfnamefont {E.~N.}\ \bibnamefont {Saridakis}},\ }\bibfield  {title} {\bibinfo {title} {{f(T) teleparallel gravity and cosmology}},\ }\href {https://doi.org/10.1088/0034-4885/79/10/106901} {\bibfield  {journal} {\bibinfo  {journal} {Rept. Prog. Phys.}\ }\textbf {\bibinfo {volume} {79}},\ \bibinfo {pages} {106901} (\bibinfo {year} {2016})},\ \Eprint {https://arxiv.org/abs/1511.07586} {arXiv:1511.07586 [gr-qc]} \BibitemShut {NoStop}%
\bibitem [{\citenamefont {Cognola}\ \emph {et~al.}(2006)\citenamefont {Cognola}, \citenamefont {Elizalde}, \citenamefont {Nojiri}, \citenamefont {Odintsov},\ and\ \citenamefont {Zerbini}}]{Cognola:2006eg}%
  \BibitemOpen
  \bibfield  {author} {\bibinfo {author} {\bibfnamefont {G.}~\bibnamefont {Cognola}}, \bibinfo {author} {\bibfnamefont {E.}~\bibnamefont {Elizalde}}, \bibinfo {author} {\bibfnamefont {S.}~\bibnamefont {Nojiri}}, \bibinfo {author} {\bibfnamefont {S.~D.}\ \bibnamefont {Odintsov}},\ and\ \bibinfo {author} {\bibfnamefont {S.}~\bibnamefont {Zerbini}},\ }\bibfield  {title} {\bibinfo {title} {{Dark energy in modified Gauss-Bonnet gravity: Late-time acceleration and the hierarchy problem}},\ }\href {https://doi.org/10.1103/PhysRevD.73.084007} {\bibfield  {journal} {\bibinfo  {journal} {Phys. Rev. D}\ }\textbf {\bibinfo {volume} {73}},\ \bibinfo {pages} {084007} (\bibinfo {year} {2006})},\ \Eprint {https://arxiv.org/abs/hep-th/0601008} {arXiv:hep-th/0601008} \BibitemShut {NoStop}%
\bibitem [{\citenamefont {Houndjo}(2012)}]{Houndjo:2011tu}%
  \BibitemOpen
  \bibfield  {author} {\bibinfo {author} {\bibfnamefont {M.~J.~S.}\ \bibnamefont {Houndjo}},\ }\bibfield  {title} {\bibinfo {title} {{Reconstruction of f(R, T) gravity describing matter dominated and accelerated phases}},\ }\href {https://doi.org/10.1142/S0218271812500034} {\bibfield  {journal} {\bibinfo  {journal} {Int. J. Mod. Phys. D}\ }\textbf {\bibinfo {volume} {21}},\ \bibinfo {pages} {1250003} (\bibinfo {year} {2012})},\ \Eprint {https://arxiv.org/abs/1107.3887} {arXiv:1107.3887 [astro-ph.CO]} \BibitemShut {NoStop}%
\bibitem [{\citenamefont {Wu}\ and\ \citenamefont {Yu}(2010)}]{Wu:2010mn}%
  \BibitemOpen
  \bibfield  {author} {\bibinfo {author} {\bibfnamefont {P.}~\bibnamefont {Wu}}\ and\ \bibinfo {author} {\bibfnamefont {H.~W.}\ \bibnamefont {Yu}},\ }\bibfield  {title} {\bibinfo {title} {{Observational constraints on $f(T)$ theory}},\ }\href {https://doi.org/10.1016/j.physletb.2010.08.073} {\bibfield  {journal} {\bibinfo  {journal} {Phys. Lett. B}\ }\textbf {\bibinfo {volume} {693}},\ \bibinfo {pages} {415} (\bibinfo {year} {2010})},\ \Eprint {https://arxiv.org/abs/1006.0674} {arXiv:1006.0674 [gr-qc]} \BibitemShut {NoStop}%
\bibitem [{\citenamefont {Harko}\ \emph {et~al.}(2018)\citenamefont {Harko}, \citenamefont {Koivisto}, \citenamefont {Lobo}, \citenamefont {Olmo},\ and\ \citenamefont {Rubiera-Garcia}}]{Harko:2018gxr}%
  \BibitemOpen
  \bibfield  {author} {\bibinfo {author} {\bibfnamefont {T.}~\bibnamefont {Harko}}, \bibinfo {author} {\bibfnamefont {T.~S.}\ \bibnamefont {Koivisto}}, \bibinfo {author} {\bibfnamefont {F.~S.~N.}\ \bibnamefont {Lobo}}, \bibinfo {author} {\bibfnamefont {G.~J.}\ \bibnamefont {Olmo}},\ and\ \bibinfo {author} {\bibfnamefont {D.}~\bibnamefont {Rubiera-Garcia}},\ }\bibfield  {title} {\bibinfo {title} {{Coupling matter in modified $Q$ gravity}},\ }\href {https://doi.org/10.1103/PhysRevD.98.084043} {\bibfield  {journal} {\bibinfo  {journal} {Phys. Rev. D}\ }\textbf {\bibinfo {volume} {98}},\ \bibinfo {pages} {084043} (\bibinfo {year} {2018})},\ \Eprint {https://arxiv.org/abs/1806.10437} {arXiv:1806.10437 [gr-qc]} \BibitemShut {NoStop}%
\bibitem [{\citenamefont {Carroll}\ \emph {et~al.}(2003)\citenamefont {Carroll}, \citenamefont {Hoffman},\ and\ \citenamefont {Trodden}}]{Carroll:2003st}%
  \BibitemOpen
  \bibfield  {author} {\bibinfo {author} {\bibfnamefont {S.~M.}\ \bibnamefont {Carroll}}, \bibinfo {author} {\bibfnamefont {M.}~\bibnamefont {Hoffman}},\ and\ \bibinfo {author} {\bibfnamefont {M.}~\bibnamefont {Trodden}},\ }\bibfield  {title} {\bibinfo {title} {{Can the dark energy equation-of-state parameter $w$ be less than $-1$?}},\ }\href {https://doi.org/10.1103/PhysRevD.68.023509} {\bibfield  {journal} {\bibinfo  {journal} {Phys. Rev. D}\ }\textbf {\bibinfo {volume} {68}},\ \bibinfo {pages} {023509} (\bibinfo {year} {2003})},\ \Eprint {https://arxiv.org/abs/astro-ph/0301273} {arXiv:astro-ph/0301273} \BibitemShut {NoStop}%
\bibitem [{\citenamefont {Padmanabhan}(2002)}]{Padmanabhan:2002cp}%
  \BibitemOpen
  \bibfield  {author} {\bibinfo {author} {\bibfnamefont {T.}~\bibnamefont {Padmanabhan}},\ }\bibfield  {title} {\bibinfo {title} {{Accelerated expansion of the universe driven by tachyonic matter}},\ }\href {https://doi.org/10.1103/PhysRevD.66.021301} {\bibfield  {journal} {\bibinfo  {journal} {Phys. Rev. D}\ }\textbf {\bibinfo {volume} {66}},\ \bibinfo {pages} {021301} (\bibinfo {year} {2002})},\ \Eprint {https://arxiv.org/abs/hep-th/0204150} {arXiv:hep-th/0204150} \BibitemShut {NoStop}%
\bibitem [{\citenamefont {Ferreira}\ and\ \citenamefont {Joyce}(1998)}]{Ferreira:1997hj}%
  \BibitemOpen
  \bibfield  {author} {\bibinfo {author} {\bibfnamefont {P.~G.}\ \bibnamefont {Ferreira}}\ and\ \bibinfo {author} {\bibfnamefont {M.}~\bibnamefont {Joyce}},\ }\bibfield  {title} {\bibinfo {title} {{Cosmology with a primordial scaling field}},\ }\href {https://doi.org/10.1103/PhysRevD.58.023503} {\bibfield  {journal} {\bibinfo  {journal} {Phys. Rev. D}\ }\textbf {\bibinfo {volume} {58}},\ \bibinfo {pages} {023503} (\bibinfo {year} {1998})},\ \Eprint {https://arxiv.org/abs/astro-ph/9711102} {arXiv:astro-ph/9711102} \BibitemShut {NoStop}%
\bibitem [{\citenamefont {Cai}\ \emph {et~al.}(2010)\citenamefont {Cai}, \citenamefont {Saridakis}, \citenamefont {Setare},\ and\ \citenamefont {Xia}}]{Cai:2009zp}%
  \BibitemOpen
  \bibfield  {author} {\bibinfo {author} {\bibfnamefont {Y.-F.}\ \bibnamefont {Cai}}, \bibinfo {author} {\bibfnamefont {E.~N.}\ \bibnamefont {Saridakis}}, \bibinfo {author} {\bibfnamefont {M.~R.}\ \bibnamefont {Setare}},\ and\ \bibinfo {author} {\bibfnamefont {J.-Q.}\ \bibnamefont {Xia}},\ }\bibfield  {title} {\bibinfo {title} {{Quintom Cosmology: Theoretical implications and observations}},\ }\href {https://doi.org/10.1016/j.physrep.2010.04.001} {\bibfield  {journal} {\bibinfo  {journal} {Phys. Rept.}\ }\textbf {\bibinfo {volume} {493}},\ \bibinfo {pages} {1} (\bibinfo {year} {2010})},\ \Eprint {https://arxiv.org/abs/0909.2776} {arXiv:0909.2776 [hep-th]} \BibitemShut {NoStop}%
\bibitem [{\citenamefont {Deffayet}\ \emph {et~al.}(2010)\citenamefont {Deffayet}, \citenamefont {Pujolas}, \citenamefont {Sawicki},\ and\ \citenamefont {Vikman}}]{Deffayet:2010qz}%
  \BibitemOpen
  \bibfield  {author} {\bibinfo {author} {\bibfnamefont {C.}~\bibnamefont {Deffayet}}, \bibinfo {author} {\bibfnamefont {O.}~\bibnamefont {Pujolas}}, \bibinfo {author} {\bibfnamefont {I.}~\bibnamefont {Sawicki}},\ and\ \bibinfo {author} {\bibfnamefont {A.}~\bibnamefont {Vikman}},\ }\bibfield  {title} {\bibinfo {title} {{Imperfect Dark Energy from Kinetic Gravity Braiding}},\ }\href {https://doi.org/10.1088/1475-7516/2010/10/026} {\bibfield  {journal} {\bibinfo  {journal} {JCAP}\ }\textbf {\bibinfo {volume} {10}},\ \bibinfo {pages} {026}},\ \Eprint {https://arxiv.org/abs/1008.0048} {arXiv:1008.0048 [hep-th]} \BibitemShut {NoStop}%
\bibitem [{\citenamefont {Zimdahl}\ and\ \citenamefont {Pavon}(2001)}]{Zimdahl:2001ar}%
  \BibitemOpen
  \bibfield  {author} {\bibinfo {author} {\bibfnamefont {W.}~\bibnamefont {Zimdahl}}\ and\ \bibinfo {author} {\bibfnamefont {D.}~\bibnamefont {Pavon}},\ }\bibfield  {title} {\bibinfo {title} {{Interacting quintessence}},\ }\href {https://doi.org/10.1016/S0370-2693(01)01174-1} {\bibfield  {journal} {\bibinfo  {journal} {Phys. Lett. B}\ }\textbf {\bibinfo {volume} {521}},\ \bibinfo {pages} {133} (\bibinfo {year} {2001})},\ \Eprint {https://arxiv.org/abs/astro-ph/0105479} {arXiv:astro-ph/0105479} \BibitemShut {NoStop}%
\bibitem [{\citenamefont {Nojiri}\ \emph {et~al.}(2005)\citenamefont {Nojiri}, \citenamefont {Odintsov},\ and\ \citenamefont {Sasaki}}]{Nojiri:2005vv}%
  \BibitemOpen
  \bibfield  {author} {\bibinfo {author} {\bibfnamefont {S.}~\bibnamefont {Nojiri}}, \bibinfo {author} {\bibfnamefont {S.~D.}\ \bibnamefont {Odintsov}},\ and\ \bibinfo {author} {\bibfnamefont {M.}~\bibnamefont {Sasaki}},\ }\bibfield  {title} {\bibinfo {title} {{Gauss-Bonnet dark energy}},\ }\href {https://doi.org/10.1103/PhysRevD.71.123509} {\bibfield  {journal} {\bibinfo  {journal} {Phys. Rev. D}\ }\textbf {\bibinfo {volume} {71}},\ \bibinfo {pages} {123509} (\bibinfo {year} {2005})},\ \Eprint {https://arxiv.org/abs/hep-th/0504052} {arXiv:hep-th/0504052} \BibitemShut {NoStop}%
\bibitem [{\citenamefont {Marciu}(2019)}]{Marciu:2019cpb}%
  \BibitemOpen
  \bibfield  {author} {\bibinfo {author} {\bibfnamefont {M.}~\bibnamefont {Marciu}},\ }\bibfield  {title} {\bibinfo {title} {{Prospects of the cosmic scenery in a quintom dark energy model with generalized nonminimal Gauss-Bonnet couplings}},\ }\href {https://doi.org/10.1103/PhysRevD.99.043508} {\bibfield  {journal} {\bibinfo  {journal} {Phys. Rev. D}\ }\textbf {\bibinfo {volume} {99}},\ \bibinfo {pages} {043508} (\bibinfo {year} {2019})}\BibitemShut {NoStop}%
\bibitem [{\citenamefont {Marciu}\ \emph {et~al.}(2018)\citenamefont {Marciu}, \citenamefont {Ioan},\ and\ \citenamefont {Iancu}}]{Marciu:2018oks}%
  \BibitemOpen
  \bibfield  {author} {\bibinfo {author} {\bibfnamefont {M.}~\bibnamefont {Marciu}}, \bibinfo {author} {\bibfnamefont {D.~M.}\ \bibnamefont {Ioan}},\ and\ \bibinfo {author} {\bibfnamefont {F.~V.}\ \bibnamefont {Iancu}},\ }\bibfield  {title} {\bibinfo {title} {{Dynamical features of a quintom dark energy model with Galileon corrections}},\ }\href {https://doi.org/10.1142/S0218271819500184} {\bibfield  {journal} {\bibinfo  {journal} {Int. J. Mod. Phys. D}\ }\textbf {\bibinfo {volume} {28}},\ \bibinfo {pages} {1950018} (\bibinfo {year} {2018})}\BibitemShut {NoStop}%
\bibitem [{\citenamefont {Bahamonde}\ \emph {et~al.}(2021)\citenamefont {Bahamonde}, \citenamefont {Marciu}, \citenamefont {Odintsov},\ and\ \citenamefont {Rudra}}]{Bahamonde:2020vfj}%
  \BibitemOpen
  \bibfield  {author} {\bibinfo {author} {\bibfnamefont {S.}~\bibnamefont {Bahamonde}}, \bibinfo {author} {\bibfnamefont {M.}~\bibnamefont {Marciu}}, \bibinfo {author} {\bibfnamefont {S.~D.}\ \bibnamefont {Odintsov}},\ and\ \bibinfo {author} {\bibfnamefont {P.}~\bibnamefont {Rudra}},\ }\bibfield  {title} {\bibinfo {title} {{String-inspired Teleparallel cosmology}},\ }\href {https://doi.org/10.1016/j.nuclphysb.2020.115238} {\bibfield  {journal} {\bibinfo  {journal} {Nucl. Phys. B}\ }\textbf {\bibinfo {volume} {962}},\ \bibinfo {pages} {115238} (\bibinfo {year} {2021})},\ \Eprint {https://arxiv.org/abs/2003.13434} {arXiv:2003.13434 [gr-qc]} \BibitemShut {NoStop}%
\bibitem [{\citenamefont {Marciu}(2020{\natexlab{a}})}]{Marciu:2020vve}%
  \BibitemOpen
  \bibfield  {author} {\bibinfo {author} {\bibfnamefont {M.}~\bibnamefont {Marciu}},\ }\bibfield  {title} {\bibinfo {title} {{Dynamical description of a quintom cosmological model nonminimally coupled with gravity}},\ }\href {https://doi.org/10.1140/epjc/s10052-020-08476-9} {\bibfield  {journal} {\bibinfo  {journal} {Eur. Phys. J. C}\ }\textbf {\bibinfo {volume} {80}},\ \bibinfo {pages} {894} (\bibinfo {year} {2020}{\natexlab{a}})},\ \Eprint {https://arxiv.org/abs/2005.03443} {arXiv:2005.03443 [gr-qc]} \BibitemShut {NoStop}%
\bibitem [{\citenamefont {Bahamonde}\ \emph {et~al.}(2018)\citenamefont {Bahamonde}, \citenamefont {B\"ohmer}, \citenamefont {Carloni}, \citenamefont {Copeland}, \citenamefont {Fang},\ and\ \citenamefont {Tamanini}}]{Bahamonde:2017ize}%
  \BibitemOpen
  \bibfield  {author} {\bibinfo {author} {\bibfnamefont {S.}~\bibnamefont {Bahamonde}}, \bibinfo {author} {\bibfnamefont {C.~G.}\ \bibnamefont {B\"ohmer}}, \bibinfo {author} {\bibfnamefont {S.}~\bibnamefont {Carloni}}, \bibinfo {author} {\bibfnamefont {E.~J.}\ \bibnamefont {Copeland}}, \bibinfo {author} {\bibfnamefont {W.}~\bibnamefont {Fang}},\ and\ \bibinfo {author} {\bibfnamefont {N.}~\bibnamefont {Tamanini}},\ }\bibfield  {title} {\bibinfo {title} {{Dynamical systems applied to cosmology: dark energy and modified gravity}},\ }\href {https://doi.org/10.1016/j.physrep.2018.09.001} {\bibfield  {journal} {\bibinfo  {journal} {Phys. Rept.}\ }\textbf {\bibinfo {volume} {775-777}},\ \bibinfo {pages} {1} (\bibinfo {year} {2018})},\ \Eprint {https://arxiv.org/abs/1712.03107} {arXiv:1712.03107 [gr-qc]} \BibitemShut {NoStop}%
\bibitem [{\citenamefont {Bueno}\ and\ \citenamefont {Cano}(2016{\natexlab{a}})}]{Bueno:2016xff}%
  \BibitemOpen
  \bibfield  {author} {\bibinfo {author} {\bibfnamefont {P.}~\bibnamefont {Bueno}}\ and\ \bibinfo {author} {\bibfnamefont {P.~A.}\ \bibnamefont {Cano}},\ }\bibfield  {title} {\bibinfo {title} {{Einsteinian cubic gravity}},\ }\href {https://doi.org/10.1103/PhysRevD.94.104005} {\bibfield  {journal} {\bibinfo  {journal} {Phys. Rev. D}\ }\textbf {\bibinfo {volume} {94}},\ \bibinfo {pages} {104005} (\bibinfo {year} {2016}{\natexlab{a}})},\ \Eprint {https://arxiv.org/abs/1607.06463} {arXiv:1607.06463 [hep-th]} \BibitemShut {NoStop}%
\bibitem [{\citenamefont {Erices}\ \emph {et~al.}(2019)\citenamefont {Erices}, \citenamefont {Papantonopoulos},\ and\ \citenamefont {Saridakis}}]{Erices:2019mkd}%
  \BibitemOpen
  \bibfield  {author} {\bibinfo {author} {\bibfnamefont {C.}~\bibnamefont {Erices}}, \bibinfo {author} {\bibfnamefont {E.}~\bibnamefont {Papantonopoulos}},\ and\ \bibinfo {author} {\bibfnamefont {E.~N.}\ \bibnamefont {Saridakis}},\ }\bibfield  {title} {\bibinfo {title} {{Cosmology in cubic and $f(P)$ gravity}},\ }\href {https://doi.org/10.1103/PhysRevD.99.123527} {\bibfield  {journal} {\bibinfo  {journal} {Phys. Rev. D}\ }\textbf {\bibinfo {volume} {99}},\ \bibinfo {pages} {123527} (\bibinfo {year} {2019})},\ \Eprint {https://arxiv.org/abs/1903.11128} {arXiv:1903.11128 [gr-qc]} \BibitemShut {NoStop}%
\bibitem [{\citenamefont {Quiros}\ \emph {et~al.}(2021)\citenamefont {Quiros}, \citenamefont {De~Arcia}, \citenamefont {Garc\'\i{}a-Salcedo}, \citenamefont {Gonzalez}, \citenamefont {Linares Cede\~no},\ and\ \citenamefont {Nucamendi}}]{Quiros:2020eim}%
  \BibitemOpen
  \bibfield  {author} {\bibinfo {author} {\bibfnamefont {I.}~\bibnamefont {Quiros}}, \bibinfo {author} {\bibfnamefont {R.}~\bibnamefont {De~Arcia}}, \bibinfo {author} {\bibfnamefont {R.}~\bibnamefont {Garc\'\i{}a-Salcedo}}, \bibinfo {author} {\bibfnamefont {T.}~\bibnamefont {Gonzalez}}, \bibinfo {author} {\bibfnamefont {F.~X.}\ \bibnamefont {Linares Cede\~no}},\ and\ \bibinfo {author} {\bibfnamefont {U.}~\bibnamefont {Nucamendi}},\ }\bibfield  {title} {\bibinfo {title} {{On the quantum origin of inflation in the geometric inflation model}},\ }\href {https://doi.org/10.1103/PhysRevD.103.064043} {\bibfield  {journal} {\bibinfo  {journal} {Phys. Rev. D}\ }\textbf {\bibinfo {volume} {103}},\ \bibinfo {pages} {064043} (\bibinfo {year} {2021})},\ \Eprint {https://arxiv.org/abs/2007.06111} {arXiv:2007.06111 [gr-qc]} \BibitemShut {NoStop}%
\bibitem [{\citenamefont {Marciu}(2020{\natexlab{b}})}]{Marciu:2020ysf}%
  \BibitemOpen
  \bibfield  {author} {\bibinfo {author} {\bibfnamefont {M.}~\bibnamefont {Marciu}},\ }\bibfield  {title} {\bibinfo {title} {{Note on the dynamical features for the extended $f(P)$ cubic gravity}},\ }\href {https://doi.org/10.1103/PhysRevD.101.103534} {\bibfield  {journal} {\bibinfo  {journal} {Phys. Rev. D}\ }\textbf {\bibinfo {volume} {101}},\ \bibinfo {pages} {103534} (\bibinfo {year} {2020}{\natexlab{b}})},\ \Eprint {https://arxiv.org/abs/2003.06403} {arXiv:2003.06403 [gr-qc]} \BibitemShut {NoStop}%
\bibitem [{\citenamefont {Quiros}\ \emph {et~al.}(2020)\citenamefont {Quiros}, \citenamefont {Garc\'\i{}a-Salcedo}, \citenamefont {Gonzalez}, \citenamefont {Mart\'\i{}nez},\ and\ \citenamefont {Nucamendi}}]{Quiros:2020uhr}%
  \BibitemOpen
  \bibfield  {author} {\bibinfo {author} {\bibfnamefont {I.}~\bibnamefont {Quiros}}, \bibinfo {author} {\bibfnamefont {R.}~\bibnamefont {Garc\'\i{}a-Salcedo}}, \bibinfo {author} {\bibfnamefont {T.}~\bibnamefont {Gonzalez}}, \bibinfo {author} {\bibfnamefont {J.~L.~M.}\ \bibnamefont {Mart\'\i{}nez}},\ and\ \bibinfo {author} {\bibfnamefont {U.}~\bibnamefont {Nucamendi}},\ }\bibfield  {title} {\bibinfo {title} {{Global asymptotic dynamics of cosmological Einsteinian cubic gravity}},\ }\href {https://doi.org/10.1103/PhysRevD.102.044018} {\bibfield  {journal} {\bibinfo  {journal} {Phys. Rev. D}\ }\textbf {\bibinfo {volume} {102}},\ \bibinfo {pages} {044018} (\bibinfo {year} {2020})},\ \Eprint {https://arxiv.org/abs/2003.10516} {arXiv:2003.10516 [gr-qc]} \BibitemShut {NoStop}%
\bibitem [{\citenamefont {Marciu}(2021)}]{Marciu:2021rdl}%
  \BibitemOpen
  \bibfield  {author} {\bibinfo {author} {\bibfnamefont {M.}~\bibnamefont {Marciu}},\ }\bibfield  {title} {\bibinfo {title} {{Dark effects in $\tilde{f}(R,P)$ gravity}},\ }\href {https://doi.org/10.1140/epjc/s10052-021-09871-6} {\bibfield  {journal} {\bibinfo  {journal} {Eur. Phys. J. C}\ }\textbf {\bibinfo {volume} {81}},\ \bibinfo {pages} {1084} (\bibinfo {year} {2021})},\ \Eprint {https://arxiv.org/abs/2103.08420} {arXiv:2103.08420 [gr-qc]} \BibitemShut {NoStop}%
\bibitem [{\citenamefont {Bueno}\ \emph {et~al.}(2017)\citenamefont {Bueno}, \citenamefont {Cano}, \citenamefont {Min},\ and\ \citenamefont {Visser}}]{Bueno:2016ypa}%
  \BibitemOpen
  \bibfield  {author} {\bibinfo {author} {\bibfnamefont {P.}~\bibnamefont {Bueno}}, \bibinfo {author} {\bibfnamefont {P.~A.}\ \bibnamefont {Cano}}, \bibinfo {author} {\bibfnamefont {V.~S.}\ \bibnamefont {Min}},\ and\ \bibinfo {author} {\bibfnamefont {M.~R.}\ \bibnamefont {Visser}},\ }\bibfield  {title} {\bibinfo {title} {{Aspects of general higher-order gravities}},\ }\href {https://doi.org/10.1103/PhysRevD.95.044010} {\bibfield  {journal} {\bibinfo  {journal} {Phys. Rev. D}\ }\textbf {\bibinfo {volume} {95}},\ \bibinfo {pages} {044010} (\bibinfo {year} {2017})},\ \Eprint {https://arxiv.org/abs/1610.08519} {arXiv:1610.08519 [hep-th]} \BibitemShut {NoStop}%
\bibitem [{\citenamefont {Edelstein}\ \emph {et~al.}(2022)\citenamefont {Edelstein}, \citenamefont {Grandi},\ and\ \citenamefont {Rivadulla~S\'anchez}}]{Edelstein:2022xlb}%
  \BibitemOpen
  \bibfield  {author} {\bibinfo {author} {\bibfnamefont {J.~D.}\ \bibnamefont {Edelstein}}, \bibinfo {author} {\bibfnamefont {N.}~\bibnamefont {Grandi}},\ and\ \bibinfo {author} {\bibfnamefont {A.}~\bibnamefont {Rivadulla~S\'anchez}},\ }\bibfield  {title} {\bibinfo {title} {{Holographic superconductivity in Einsteinian Cubic Gravity}},\ }\href {https://doi.org/10.1007/JHEP05(2022)188} {\bibfield  {journal} {\bibinfo  {journal} {JHEP}\ }\textbf {\bibinfo {volume} {05}},\ \bibinfo {pages} {188}},\ \Eprint {https://arxiv.org/abs/2202.05781} {arXiv:2202.05781 [hep-th]} \BibitemShut {NoStop}%
\bibitem [{\citenamefont {Bueno}\ \emph {et~al.}(2019)\citenamefont {Bueno}, \citenamefont {Cano}, \citenamefont {Moreno},\ and\ \citenamefont {Murcia}}]{Bueno:2019ltp}%
  \BibitemOpen
  \bibfield  {author} {\bibinfo {author} {\bibfnamefont {P.}~\bibnamefont {Bueno}}, \bibinfo {author} {\bibfnamefont {P.~A.}\ \bibnamefont {Cano}}, \bibinfo {author} {\bibfnamefont {J.}~\bibnamefont {Moreno}},\ and\ \bibinfo {author} {\bibfnamefont {A.}~\bibnamefont {Murcia}},\ }\bibfield  {title} {\bibinfo {title} {{All higher-curvature gravities as Generalized quasi-topological gravities}},\ }\href {https://doi.org/10.1007/JHEP11(2019)062} {\bibfield  {journal} {\bibinfo  {journal} {JHEP}\ }\textbf {\bibinfo {volume} {11}},\ \bibinfo {pages} {062}},\ \Eprint {https://arxiv.org/abs/1906.00987} {arXiv:1906.00987 [hep-th]} \BibitemShut {NoStop}%
\bibitem [{\citenamefont {C\'aceres}\ \emph {et~al.}(2021)\citenamefont {C\'aceres}, \citenamefont {V\'asquez},\ and\ \citenamefont {Vilar~L\'opez}}]{Caceres:2020jrf}%
  \BibitemOpen
  \bibfield  {author} {\bibinfo {author} {\bibfnamefont {E.}~\bibnamefont {C\'aceres}}, \bibinfo {author} {\bibfnamefont {R.~C.}\ \bibnamefont {V\'asquez}},\ and\ \bibinfo {author} {\bibfnamefont {A.}~\bibnamefont {Vilar~L\'opez}},\ }\bibfield  {title} {\bibinfo {title} {{Entanglement entropy in cubic gravitational theories}},\ }\href {https://doi.org/10.1007/JHEP05(2021)186} {\bibfield  {journal} {\bibinfo  {journal} {JHEP}\ }\textbf {\bibinfo {volume} {05}},\ \bibinfo {pages} {186}},\ \Eprint {https://arxiv.org/abs/2009.11595} {arXiv:2009.11595 [hep-th]} \BibitemShut {NoStop}%
\bibitem [{\citenamefont {Rudra}(2023)}]{Rudra:2022qbv}%
  \BibitemOpen
  \bibfield  {author} {\bibinfo {author} {\bibfnamefont {P.}~\bibnamefont {Rudra}},\ }\bibfield  {title} {\bibinfo {title} {{Ricci-cubic holographic dark energy}},\ }\href {https://doi.org/10.1016/j.dark.2023.101307} {\bibfield  {journal} {\bibinfo  {journal} {Phys. Dark Univ.}\ }\textbf {\bibinfo {volume} {42}},\ \bibinfo {pages} {101307} (\bibinfo {year} {2023})},\ \Eprint {https://arxiv.org/abs/2206.03490} {arXiv:2206.03490 [gr-qc]} \BibitemShut {NoStop}%
\bibitem [{\citenamefont {Bueno}\ \emph {et~al.}(2018)\citenamefont {Bueno}, \citenamefont {Cano},\ and\ \citenamefont {Ruip\'erez}}]{Bueno:2018xqc}%
  \BibitemOpen
  \bibfield  {author} {\bibinfo {author} {\bibfnamefont {P.}~\bibnamefont {Bueno}}, \bibinfo {author} {\bibfnamefont {P.~A.}\ \bibnamefont {Cano}},\ and\ \bibinfo {author} {\bibfnamefont {A.}~\bibnamefont {Ruip\'erez}},\ }\bibfield  {title} {\bibinfo {title} {{Holographic studies of Einsteinian cubic gravity}},\ }\href {https://doi.org/10.1007/JHEP03(2018)150} {\bibfield  {journal} {\bibinfo  {journal} {JHEP}\ }\textbf {\bibinfo {volume} {03}},\ \bibinfo {pages} {150}},\ \Eprint {https://arxiv.org/abs/1802.00018} {arXiv:1802.00018 [hep-th]} \BibitemShut {NoStop}%
\bibitem [{\citenamefont {Beltr\'an~Jim\'enez}\ and\ \citenamefont {Jim\'enez-Cano}(2021)}]{BeltranJimenez:2020lee}%
  \BibitemOpen
  \bibfield  {author} {\bibinfo {author} {\bibfnamefont {J.}~\bibnamefont {Beltr\'an~Jim\'enez}}\ and\ \bibinfo {author} {\bibfnamefont {A.}~\bibnamefont {Jim\'enez-Cano}},\ }\bibfield  {title} {\bibinfo {title} {{On the strong coupling of Einsteinian Cubic Gravity and its generalisations}},\ }\href {https://doi.org/10.1088/1475-7516/2021/01/069} {\bibfield  {journal} {\bibinfo  {journal} {JCAP}\ }\textbf {\bibinfo {volume} {01}},\ \bibinfo {pages} {069}},\ \Eprint {https://arxiv.org/abs/2009.08197} {arXiv:2009.08197 [gr-qc]} \BibitemShut {NoStop}%
\bibitem [{\citenamefont {Giri}\ and\ \citenamefont {Rudra}(2022)}]{Giri:2021amc}%
  \BibitemOpen
  \bibfield  {author} {\bibinfo {author} {\bibfnamefont {K.}~\bibnamefont {Giri}}\ and\ \bibinfo {author} {\bibfnamefont {P.}~\bibnamefont {Rudra}},\ }\bibfield  {title} {\bibinfo {title} {{Constraints on cubic and f(P) gravity from the cosmic chronometers, BAO \& CMB datasets: Use of machine learning algorithms}},\ }\href {https://doi.org/10.1016/j.nuclphysb.2022.115746} {\bibfield  {journal} {\bibinfo  {journal} {Nucl. Phys. B}\ }\textbf {\bibinfo {volume} {978}},\ \bibinfo {pages} {115746} (\bibinfo {year} {2022})},\ \Eprint {https://arxiv.org/abs/2107.12417} {arXiv:2107.12417 [astro-ph.CO]} \BibitemShut {NoStop}%
\bibitem [{\citenamefont {Marciu}(2022)}]{Marciu:2022wzh}%
  \BibitemOpen
  \bibfield  {author} {\bibinfo {author} {\bibfnamefont {M.}~\bibnamefont {Marciu}},\ }\bibfield  {title} {\bibinfo {title} {{Tachyonic cosmology with cubic contractions of the Riemann tensor}},\ }\href {https://doi.org/10.1140/epjc/s10052-022-11023-3} {\bibfield  {journal} {\bibinfo  {journal} {Eur. Phys. J. C}\ }\textbf {\bibinfo {volume} {82}},\ \bibinfo {pages} {1069} (\bibinfo {year} {2022})},\ \Eprint {https://arxiv.org/abs/2203.00598} {arXiv:2203.00598 [gr-qc]} \BibitemShut {NoStop}%
\bibitem [{\citenamefont {Marciu}(2020{\natexlab{c}})}]{Marciu:2020ski}%
  \BibitemOpen
  \bibfield  {author} {\bibinfo {author} {\bibfnamefont {M.}~\bibnamefont {Marciu}},\ }\bibfield  {title} {\bibinfo {title} {{Dynamical aspects for scalar fields coupled to cubic contractions of the Riemann tensor}},\ }\href {https://doi.org/10.1103/PhysRevD.102.023517} {\bibfield  {journal} {\bibinfo  {journal} {Phys. Rev. D}\ }\textbf {\bibinfo {volume} {102}},\ \bibinfo {pages} {023517} (\bibinfo {year} {2020}{\natexlab{c}})},\ \Eprint {https://arxiv.org/abs/2004.07120} {arXiv:2004.07120 [gr-qc]} \BibitemShut {NoStop}%
\bibitem [{\citenamefont {Marciu}(2023)}]{Marciu2023}%
  \BibitemOpen
  \bibfield  {author} {\bibinfo {author} {\bibfnamefont {M.}~\bibnamefont {Marciu}},\ }\bibfield  {title} {\bibinfo {title} {A two-field dark energy model with cubic contractions of the riemann tensor},\ }\href {https://doi.org/10.1139/cjp-2022-0321} {\bibfield  {journal} {\bibinfo  {journal} {Canadian Journal of Physics}\ }\textbf {\bibinfo {volume} {101}},\ \bibinfo {pages} {460} (\bibinfo {year} {2023})}\BibitemShut {NoStop}%
\bibitem [{\citenamefont {Arciniega}\ \emph {et~al.}(2020{\natexlab{a}})\citenamefont {Arciniega}, \citenamefont {Edelstein},\ and\ \citenamefont {Jaime}}]{Arciniega:2018fxj}%
  \BibitemOpen
  \bibfield  {author} {\bibinfo {author} {\bibfnamefont {G.}~\bibnamefont {Arciniega}}, \bibinfo {author} {\bibfnamefont {J.~D.}\ \bibnamefont {Edelstein}},\ and\ \bibinfo {author} {\bibfnamefont {L.~G.}\ \bibnamefont {Jaime}},\ }\bibfield  {title} {\bibinfo {title} {{Towards geometric inflation: the cubic case}},\ }\href {https://doi.org/10.1016/j.physletb.2020.135272} {\bibfield  {journal} {\bibinfo  {journal} {Phys. Lett. B}\ }\textbf {\bibinfo {volume} {802}},\ \bibinfo {pages} {135272} (\bibinfo {year} {2020}{\natexlab{a}})},\ \Eprint {https://arxiv.org/abs/1810.08166} {arXiv:1810.08166 [gr-qc]} \BibitemShut {NoStop}%
\bibitem [{\citenamefont {Edelstein}\ \emph {et~al.}(2020)\citenamefont {Edelstein}, \citenamefont {V\'azquez~Rodr\'\i{}guez},\ and\ \citenamefont {Vilar~L\'opez}}]{Edelstein:2020nhg}%
  \BibitemOpen
  \bibfield  {author} {\bibinfo {author} {\bibfnamefont {J.~D.}\ \bibnamefont {Edelstein}}, \bibinfo {author} {\bibfnamefont {D.}~\bibnamefont {V\'azquez~Rodr\'\i{}guez}},\ and\ \bibinfo {author} {\bibfnamefont {A.}~\bibnamefont {Vilar~L\'opez}},\ }\bibfield  {title} {\bibinfo {title} {{Aspects of Geometric Inflation}},\ }\href {https://doi.org/10.1088/1475-7516/2020/12/040} {\bibfield  {journal} {\bibinfo  {journal} {JCAP}\ }\textbf {\bibinfo {volume} {12}},\ \bibinfo {pages} {040}},\ \Eprint {https://arxiv.org/abs/2006.10007} {arXiv:2006.10007 [hep-th]} \BibitemShut {NoStop}%
\bibitem [{\citenamefont {Arciniega}\ \emph {et~al.}(2019)\citenamefont {Arciniega}, \citenamefont {Bueno}, \citenamefont {Cano}, \citenamefont {Edelstein}, \citenamefont {Hennigar},\ and\ \citenamefont {Jaime}}]{Arciniega:2019oxa}%
  \BibitemOpen
  \bibfield  {author} {\bibinfo {author} {\bibfnamefont {G.}~\bibnamefont {Arciniega}}, \bibinfo {author} {\bibfnamefont {P.}~\bibnamefont {Bueno}}, \bibinfo {author} {\bibfnamefont {P.~A.}\ \bibnamefont {Cano}}, \bibinfo {author} {\bibfnamefont {J.~D.}\ \bibnamefont {Edelstein}}, \bibinfo {author} {\bibfnamefont {R.~A.}\ \bibnamefont {Hennigar}},\ and\ \bibinfo {author} {\bibfnamefont {L.~G.}\ \bibnamefont {Jaime}},\ }\bibfield  {title} {\bibinfo {title} {{Cosmic inflation without inflaton}},\ }\href {https://doi.org/10.1142/S0218271819440085} {\bibfield  {journal} {\bibinfo  {journal} {Int. J. Mod. Phys. D}\ }\textbf {\bibinfo {volume} {28}},\ \bibinfo {pages} {1944008} (\bibinfo {year} {2019})}\BibitemShut {NoStop}%
\bibitem [{\citenamefont {Arciniega}\ \emph {et~al.}(2020{\natexlab{b}})\citenamefont {Arciniega}, \citenamefont {Bueno}, \citenamefont {Cano}, \citenamefont {Edelstein}, \citenamefont {Hennigar},\ and\ \citenamefont {Jaime}}]{Arciniega:2018tnn}%
  \BibitemOpen
  \bibfield  {author} {\bibinfo {author} {\bibfnamefont {G.}~\bibnamefont {Arciniega}}, \bibinfo {author} {\bibfnamefont {P.}~\bibnamefont {Bueno}}, \bibinfo {author} {\bibfnamefont {P.~A.}\ \bibnamefont {Cano}}, \bibinfo {author} {\bibfnamefont {J.~D.}\ \bibnamefont {Edelstein}}, \bibinfo {author} {\bibfnamefont {R.~A.}\ \bibnamefont {Hennigar}},\ and\ \bibinfo {author} {\bibfnamefont {L.~G.}\ \bibnamefont {Jaime}},\ }\bibfield  {title} {\bibinfo {title} {{Geometric Inflation}},\ }\href {https://doi.org/10.1016/j.physletb.2020.135242} {\bibfield  {journal} {\bibinfo  {journal} {Phys. Lett. B}\ }\textbf {\bibinfo {volume} {802}},\ \bibinfo {pages} {135242} (\bibinfo {year} {2020}{\natexlab{b}})},\ \Eprint {https://arxiv.org/abs/1812.11187} {arXiv:1812.11187 [hep-th]} \BibitemShut {NoStop}%
\bibitem [{\citenamefont {Adair}\ \emph {et~al.}(2020)\citenamefont {Adair}, \citenamefont {Bueno}, \citenamefont {Cano}, \citenamefont {Hennigar},\ and\ \citenamefont {Mann}}]{Adair:2020vso}%
  \BibitemOpen
  \bibfield  {author} {\bibinfo {author} {\bibfnamefont {C.}~\bibnamefont {Adair}}, \bibinfo {author} {\bibfnamefont {P.}~\bibnamefont {Bueno}}, \bibinfo {author} {\bibfnamefont {P.~A.}\ \bibnamefont {Cano}}, \bibinfo {author} {\bibfnamefont {R.~A.}\ \bibnamefont {Hennigar}},\ and\ \bibinfo {author} {\bibfnamefont {R.~B.}\ \bibnamefont {Mann}},\ }\bibfield  {title} {\bibinfo {title} {{Slowly rotating black holes in Einsteinian cubic gravity}},\ }\href {https://doi.org/10.1103/PhysRevD.102.084001} {\bibfield  {journal} {\bibinfo  {journal} {Phys. Rev. D}\ }\textbf {\bibinfo {volume} {102}},\ \bibinfo {pages} {084001} (\bibinfo {year} {2020})},\ \Eprint {https://arxiv.org/abs/2004.09598} {arXiv:2004.09598 [gr-qc]} \BibitemShut {NoStop}%
\bibitem [{\citenamefont {Bueno}\ and\ \citenamefont {Cano}(2016{\natexlab{b}})}]{Bueno:2016lrh}%
  \BibitemOpen
  \bibfield  {author} {\bibinfo {author} {\bibfnamefont {P.}~\bibnamefont {Bueno}}\ and\ \bibinfo {author} {\bibfnamefont {P.~A.}\ \bibnamefont {Cano}},\ }\bibfield  {title} {\bibinfo {title} {{Four-dimensional black holes in Einsteinian cubic gravity}},\ }\href {https://doi.org/10.1103/PhysRevD.94.124051} {\bibfield  {journal} {\bibinfo  {journal} {Phys. Rev. D}\ }\textbf {\bibinfo {volume} {94}},\ \bibinfo {pages} {124051} (\bibinfo {year} {2016}{\natexlab{b}})},\ \Eprint {https://arxiv.org/abs/1610.08019} {arXiv:1610.08019 [hep-th]} \BibitemShut {NoStop}%
\bibitem [{\citenamefont {Hennigar}\ and\ \citenamefont {Mann}(2017)}]{Hennigar:2016gkm}%
  \BibitemOpen
  \bibfield  {author} {\bibinfo {author} {\bibfnamefont {R.~A.}\ \bibnamefont {Hennigar}}\ and\ \bibinfo {author} {\bibfnamefont {R.~B.}\ \bibnamefont {Mann}},\ }\bibfield  {title} {\bibinfo {title} {{Black holes in Einsteinian cubic gravity}},\ }\href {https://doi.org/10.1103/PhysRevD.95.064055} {\bibfield  {journal} {\bibinfo  {journal} {Phys. Rev. D}\ }\textbf {\bibinfo {volume} {95}},\ \bibinfo {pages} {064055} (\bibinfo {year} {2017})},\ \Eprint {https://arxiv.org/abs/1610.06675} {arXiv:1610.06675 [hep-th]} \BibitemShut {NoStop}%
\bibitem [{\citenamefont {Feng}\ \emph {et~al.}(2017)\citenamefont {Feng}, \citenamefont {Huang}, \citenamefont {Mai},\ and\ \citenamefont {Lu}}]{Feng:2017tev}%
  \BibitemOpen
  \bibfield  {author} {\bibinfo {author} {\bibfnamefont {X.-H.}\ \bibnamefont {Feng}}, \bibinfo {author} {\bibfnamefont {H.}~\bibnamefont {Huang}}, \bibinfo {author} {\bibfnamefont {Z.-F.}\ \bibnamefont {Mai}},\ and\ \bibinfo {author} {\bibfnamefont {H.}~\bibnamefont {Lu}},\ }\bibfield  {title} {\bibinfo {title} {{Bounce Universe and Black Holes from Critical Einsteinian Cubic Gravity}},\ }\href {https://doi.org/10.1103/PhysRevD.96.104034} {\bibfield  {journal} {\bibinfo  {journal} {Phys. Rev. D}\ }\textbf {\bibinfo {volume} {96}},\ \bibinfo {pages} {104034} (\bibinfo {year} {2017})},\ \Eprint {https://arxiv.org/abs/1707.06308} {arXiv:1707.06308 [hep-th]} \BibitemShut {NoStop}%
\bibitem [{\citenamefont {Pookkillath}\ \emph {et~al.}(2020)\citenamefont {Pookkillath}, \citenamefont {De~Felice},\ and\ \citenamefont {Starobinsky}}]{Pookkillath:2020iqq}%
  \BibitemOpen
  \bibfield  {author} {\bibinfo {author} {\bibfnamefont {M.~C.}\ \bibnamefont {Pookkillath}}, \bibinfo {author} {\bibfnamefont {A.}~\bibnamefont {De~Felice}},\ and\ \bibinfo {author} {\bibfnamefont {A.~A.}\ \bibnamefont {Starobinsky}},\ }\bibfield  {title} {\bibinfo {title} {{Anisotropic instability in a higher order gravity theory}},\ }\href {https://doi.org/10.1088/1475-7516/2020/07/041} {\bibfield  {journal} {\bibinfo  {journal} {JCAP}\ }\textbf {\bibinfo {volume} {07}},\ \bibinfo {pages} {041}},\ \Eprint {https://arxiv.org/abs/2004.03912} {arXiv:2004.03912 [gr-qc]} \BibitemShut {NoStop}%
\bibitem [{\citenamefont {Hennigar}\ \emph {et~al.}(2018)\citenamefont {Hennigar}, \citenamefont {Poshteh},\ and\ \citenamefont {Mann}}]{Hennigar:2018hza}%
  \BibitemOpen
  \bibfield  {author} {\bibinfo {author} {\bibfnamefont {R.~A.}\ \bibnamefont {Hennigar}}, \bibinfo {author} {\bibfnamefont {M.~B.~J.}\ \bibnamefont {Poshteh}},\ and\ \bibinfo {author} {\bibfnamefont {R.~B.}\ \bibnamefont {Mann}},\ }\bibfield  {title} {\bibinfo {title} {{Shadows, Signals, and Stability in Einsteinian Cubic Gravity}},\ }\href {https://doi.org/10.1103/PhysRevD.97.064041} {\bibfield  {journal} {\bibinfo  {journal} {Phys. Rev. D}\ }\textbf {\bibinfo {volume} {97}},\ \bibinfo {pages} {064041} (\bibinfo {year} {2018})},\ \Eprint {https://arxiv.org/abs/1801.03223} {arXiv:1801.03223 [gr-qc]} \BibitemShut {NoStop}%
\bibitem [{\citenamefont {Sardar}\ and\ \citenamefont {Debnath}(2022)}]{Sardar:2021blt}%
  \BibitemOpen
  \bibfield  {author} {\bibinfo {author} {\bibfnamefont {A.}~\bibnamefont {Sardar}}\ and\ \bibinfo {author} {\bibfnamefont {U.}~\bibnamefont {Debnath}},\ }\bibfield  {title} {\bibinfo {title} {{Reconstruction of extended f(P) cubic gravity from other modified gravity models}},\ }\href {https://doi.org/10.1016/j.dark.2021.100926} {\bibfield  {journal} {\bibinfo  {journal} {Phys. Dark Univ.}\ }\textbf {\bibinfo {volume} {35}},\ \bibinfo {pages} {100926} (\bibinfo {year} {2022})}\BibitemShut {NoStop}%
\bibitem [{\citenamefont {Bernardo}\ \emph {et~al.}(2022)\citenamefont {Bernardo}, \citenamefont {Chen}, \citenamefont {Said~Levi},\ and\ \citenamefont {Kung}}]{Bernardo:2021ynf}%
  \BibitemOpen
  \bibfield  {author} {\bibinfo {author} {\bibfnamefont {R.~C.}\ \bibnamefont {Bernardo}}, \bibinfo {author} {\bibfnamefont {C.-Y.}\ \bibnamefont {Chen}}, \bibinfo {author} {\bibfnamefont {J.}~\bibnamefont {Said~Levi}},\ and\ \bibinfo {author} {\bibfnamefont {Y.-H.}\ \bibnamefont {Kung}},\ }\bibfield  {title} {\bibinfo {title} {{Confronting quantum-corrected teleparallel cosmology with observations}},\ }\href {https://doi.org/10.1088/1475-7516/2022/04/052} {\bibfield  {journal} {\bibinfo  {journal} {JCAP}\ }\textbf {\bibinfo {volume} {04}}\bibfield  {number} {\bibinfo  {number} { (04)},\ \bibinfo {pages} {052}},\ }\Eprint {https://arxiv.org/abs/2111.11761} {arXiv:2111.11761 [gr-qc]} \BibitemShut {NoStop}%
\bibitem [{\citenamefont {Torrado}\ and\ \citenamefont {Lewis}(2021)}]{Torrado:2020dgo}%
  \BibitemOpen
  \bibfield  {author} {\bibinfo {author} {\bibfnamefont {J.}~\bibnamefont {Torrado}}\ and\ \bibinfo {author} {\bibfnamefont {A.}~\bibnamefont {Lewis}},\ }\bibfield  {title} {\bibinfo {title} {{Cobaya: Code for Bayesian Analysis of hierarchical physical models}},\ }\href {https://doi.org/10.1088/1475-7516/2021/05/057} {\bibfield  {journal} {\bibinfo  {journal} {JCAP}\ }\textbf {\bibinfo {volume} {05}},\ \bibinfo {pages} {057}},\ \Eprint {https://arxiv.org/abs/2005.05290} {arXiv:2005.05290 [astro-ph.IM]} \BibitemShut {NoStop}%
\bibitem [{\citenamefont {Scolnic}\ \emph {et~al.}(2018)\citenamefont {Scolnic} \emph {et~al.}}]{Pan-STARRS1:2017jku}%
  \BibitemOpen
  \bibfield  {author} {\bibinfo {author} {\bibfnamefont {D.~M.}\ \bibnamefont {Scolnic}} \emph {et~al.} (\bibinfo {collaboration} {Pan-STARRS1}),\ }\bibfield  {title} {\bibinfo {title} {{The Complete Light-curve Sample of Spectroscopically Confirmed SNe Ia from Pan-STARRS1 and Cosmological Constraints from the Combined Pantheon Sample}},\ }\href {https://doi.org/10.3847/1538-4357/aab9bb} {\bibfield  {journal} {\bibinfo  {journal} {Astrophys. J.}\ }\textbf {\bibinfo {volume} {859}},\ \bibinfo {pages} {101} (\bibinfo {year} {2018})},\ \Eprint {https://arxiv.org/abs/1710.00845} {arXiv:1710.00845 [astro-ph.CO]} \BibitemShut {NoStop}%
\bibitem [{\citenamefont {{Wolfram Research{,} https://www.wolfram.com/mathematica}}()}]{Mathematica}%
  \BibitemOpen
  \bibfield  {author} {\bibinfo {author} {\bibnamefont {{Wolfram Research{,} https://www.wolfram.com/mathematica}}},\ }\href@noop {} {\bibinfo {title} {{Mathematica}}}\BibitemShut {NoStop}%
\bibitem [{\citenamefont {{J. M. Martin-Garcia}}()}]{xact}%
  \BibitemOpen
  \bibfield  {author} {\bibinfo {author} {\bibnamefont {{J. M. Martin-Garcia}}},\ }\href {http://www.xact.es} {\bibinfo {title} {{xAct: Efficient tensor computer algebra for the Wolfram Language}}}\BibitemShut {NoStop}%
\end{thebibliography}%

\end{document}